%% file: main.tex
\documentclass[cameraready]{Interspeech}

\title{Robocalls: A Worldwide or US-only Problem?\\Analyzing Spam and Fraud in International Phone Calls}

\author[orcid=0000-0002-7313-2800]{Kemal}{Altwlkany}
\author[orcid=0000-0003-1229-3149]{Andro}{Merćep}
\author[orcid=0000-0002-7229-9596]{Tomislav}{Đuričić}
\author[orcid=0000-0001-5507-5788]{Ante}{Kapetanović}
\author[orcid=0000-0002-3059-0502]{Emanuel}{Lacic}

\address{
    Infobip
}

\email{[name.surname]@infobip.com}


\usepackage[T1]{fontenc}
\usepackage[utf8]{inputenc} % needed for pdfLaTeX
\usepackage{comment}
\usepackage{subcaption}
\usepackage{csquotes}
\usepackage{amsmath}

\usepackage{xurl}

\usepackage{tabularx}
\usepackage{array} 

\newcolumntype{Y}{>{\raggedright\arraybackslash}X}

\newcommand{\USD}[1]{\$#1}

\begin{document}

\bstctlcite{bstctl:nodash}  % IEEEtran - turn off dashes for repeated authors

\maketitle

\begin{abstract}
    Unsolicited automated phone calls (robocalls) are a serious threat: in the US alone, these calls resulted in reported losses of \USD{1.1} billion during 2025. Phishing and spoofing consistently rank among the most reported crimes within the FBI's Internet Crime Complaint Center, with phone call scams having the highest reported median loss.
    Combating robocalls is difficult due to many legal and practical constraints: robocalls often encompass multiple legal jurisdictions of different countries/states, the large volume of robocalls, their multilingual nature, the lack of publicly available data, privacy concerns with obtaining data, etc.
    We present a study of international robocalls, aggregating robocall reports from countries across all inhabited continents and contribute by providing new findings on international robocalls from 65 different countries.
    We also present the first publicly available multimodal and international robocall dataset\footnote{Available here: 10.5281/zenodo.21066049}: 8.7 million call detail records, 839 robocall transcripts from 28 identified robocall campaign clusters, and 677 robocall recordings.
    We describe our methodology for collecting robocall data over a 9-month period and provide a detailed analysis comparing robocalls in the US with those in other countries.
    Our analysis covers several aspects, including uncovering calling patterns, identifying co-targeting attacks, discovering common robocall campaigns, extracting callback numbers, analyzing linguistic differences among robocalls in the same language but different regions, and other insights.
    Our results indicate that although robocalls are an international problem, the severity of the threat is significantly higher in the US than in other countries.
    We provide steps for future research and suggest remedies to reduce the effectiveness of robocalls based on our analysis.
\end{abstract}

\input{text/introduction}
\input{text/background}

\input{text/related_work}

\input{text/method}

\input{text/metadata_analysis}

\input{text/content_analysis}

\input{text/discussion}
\input{text/conclusion}

\bibliographystyle{IEEEtran}
\bibliography{mybib}

\end{document}

%% file: text/introduction.tex
\section{Introduction}

\begin{figure}[t]
  \centering
  \includegraphics[width=1.0\linewidth]{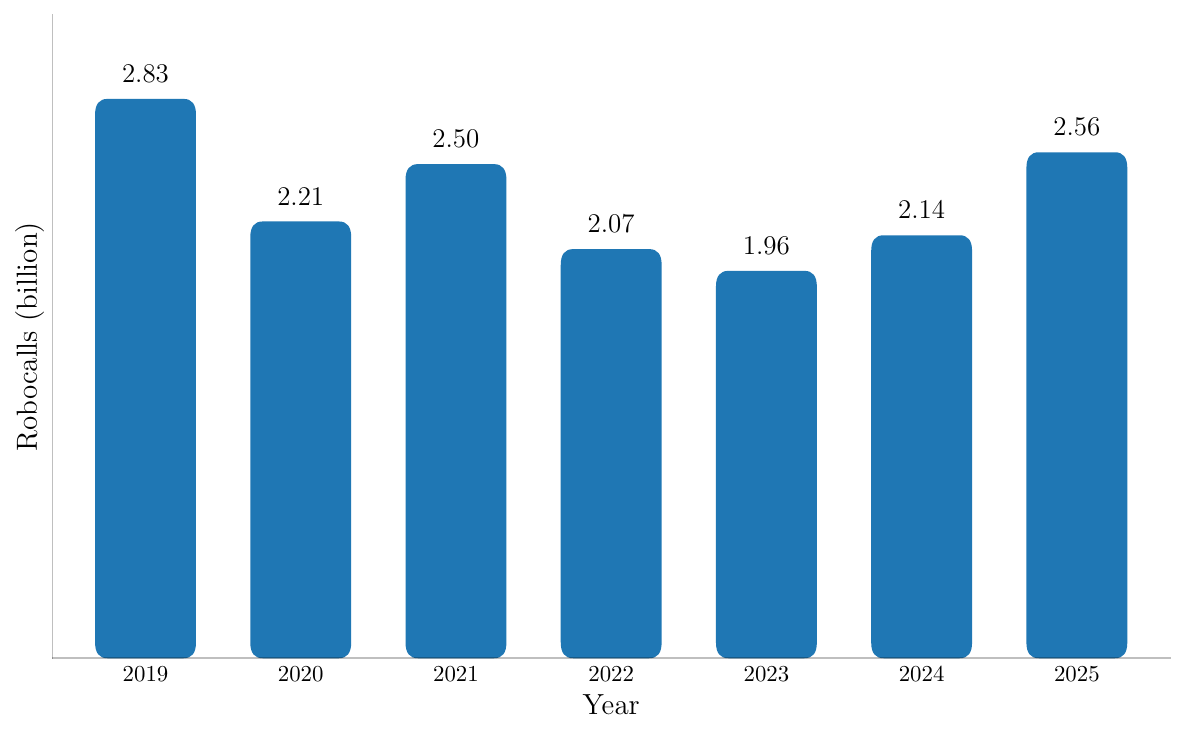}
  \caption{The average number of monthly robocalls in the US over 2025 was 2.56 billion, which is the highest since 2019.}
  \label{fig:number-of-robocalls-US}
\end{figure}

Fifteen years ago, most readers would probably have been unfamiliar with the term \textit{robocall}, and the broader public would likely never have personally encountered a robocall. This is no longer the case as many people worldwide, especially in the US, are continuously exposed to robocalls, and the problem shows no signs of slowing down. In \cref{fig:number-of-robocalls-US}, we show the average number of monthly reported robocalls for the previous 7 years (from 2019 to 2025). These numbers were obtained from the US Public Interest Research Group (PIRG) Education Fund~\cite{pirg2025ringing} and, as can be observed, robocalls have been on the rise over the past three years and reached a six-year high in 2025. In December 2025 alone, a total of 4.1 billion robocalls were made to US consumers, with the volume of scam and telemarketing calls increasing by 15.6\% in 2025 compared to 2024~\cite{pirg2026phonescams}.

\subsection{What exactly are robocalls?}

Definitions vary. According to the US Federal Communications Commission (FCC): ``\textit{robocalls are calls made with an autodialer or that contain a prerecorded or artificial voice message}''~\cite{fcc2025robocalls}. In~\cite{prasad2020s}, the authors define robocalls as a \textit{catch-all term for automated or semi-automated unsolicited calls, often for fraud or telemarketing purposes}. Some authors refer to robocalls as ``\textit{computer-generated voices}''~\cite{elizalde2021detection}. 

Not all robocalls are necessarily malicious or illegal. Some are even highly desirable, such as public safety announcements or school closure announcements~\cite{fcc2025robocalls,prasad2020s}. In the US, telemarketers are allowed to make calls using prerecorded or artificial voices, provided that they have obtained prior express written consent from the callee and that these calls are made between \DTMtime{8:00:00} and \DTMtime{21:00:00}~\cite{fcc2025robocalls}. The FCC's website contains more information regarding the legality of these calls in the US.

Robocalls are not a minor inconvenience; in fact, illegal robocalls are the FCC's top consumer complaint and top consumer protection priority~\cite{fcc2025robocalls}. According to the US Federal Trade Commission (FTC), in 2025, US citizens have reported a total of \USD{1.1} billion lost to fraud over phone calls~\cite{ftc2025fraudreports}. Therefore, stopping robocalls is of interest to the general public.

Given that the two definitions of robocalls provided by the FCC~\cite{fcc2025robocalls} and by Prasad et al.~\cite{prasad2020s} are highly aligned, in this paper, we will consider any unsolicited calls that contain a prerecorded or artificial voice message as robocalls. Such calls do not necessarily have to be malicious. We will provide more details and a stricter terminology in the following section.

\subsection{Why now?}

There are several reasons why robocalls are an ever-growing issue, but they all more or less point to the fact that it is easier than ever to create robocalls and that it is difficult to trace them, thereby making legal prosecution difficult :

\begin{itemize}
    \item The telephone network is still to date (2025) the largest network of human users, with more active users than the internet and more mobile-cellular subscriptions than the Earth's population~\cite{itu2024facts,itu2025facts}. This means that malicious users have a large pool of targets to choose from.
    \item It is difficult to regulate and trace malicious calls, as calls can be made from any place on Earth while targeting users worldwide. This makes legal action more difficult as it requires international cooperation, moreover, robocalls are regulated differently depending on the laws and regulations of each country. The low chance of being caught makes it appealing for malicious users to attempt fraudulent behavior.
    \item The telephone network operates in such a manner that the actual originating number does not have to match the caller ID which is displayed on the callee's device. This is known as \textit{caller ID spoofing} and will be discussed in more detail in the paper.
    \item The rapid advancements in technology, most notably AI, have made it very easy to generate artificial voices that sound very human-like~\cite{cui-etal-2025-recent}. Text-to-speech technology and translation tools enable malicious users to synthesize audio in languages they do not even understand.
    \item Similarly to the previous point, advancements in technology, such as coding/programming assistants~\cite{jiang2024survey}, now enable users with no technical/programming expertise to interact with technology, specifically software and APIs, that require(d) programming knowledge. This makes it easy for them to leverage cloud communication providers for malicious acts.
    \item Creating calls is cheap, especially since unanswered calls are not charged at all.
    \item Finally, such fraudulent behavior has a track record of being successful, as millions have already been stolen from people, making the misbehavior appealing to malicious users~\cite{fcc2025robocalls,ftc2025fraudreports}.
\end{itemize}

\subsection{Research questions}
This paper addresses the following research questions:
\begin{itemize}
    \item \textbf{RQ1}: Are robocalls a US problem only, or is there an international threat posed by robocalls?
    \item \textbf{RQ2}: Is there a difference between robocalls directed towards the US and other countries?
    \item \textbf{RQ3}: Uncovering robocall patterns, such as:
    \begin{itemize}
        \item What are the most common types of scam?
        \item What languages are used by robocallers?
        \item Are there temporal patterns to robocalls?
        \item Do robocallers operate in coordinated attacks?
        \item Is there evidence of robocallers using speech technology to determine if they are being monitored?
    \end{itemize}
\end{itemize}

\subsection{Paper contributions}

Among others, the main contributions of our paper can be summarized as:
\begin{itemize}
    \item We provide the first international and multilingual analysis of robocalls. To the best of our knowledge, existing literature on the topic of robocall analysis mostly addresses US and North American data, with no available analysis of robocalls within Africa, Asia, Australia, Europe, or South America. Naturally, this means that existing literature focuses on robocalls that are mostly in English language. We also provide a detailed investigation of telephony-related fraud in several countries worldwide and report on measures taken to combat fraud in telephony.
    \item We set up a honeypot consisting of more than 100 thousand phone numbers with the intent of recording robocall activity. These phone numbers are international, and, to the best of our knowledge, represent the first international study of robocall activity.
    \item We publish the first open dataset containing more than 8.7 million call detail records, recorded by our honeypot. We are not aware of any prior open datasets of call detail records.
    \item We contribute by making available 677 robocall audio recordings and 839 robocall transcripts. These recordings and transcripts are in:  English, Spanish, Chinese, and Polish and have been manually verified by at least one human annotator to be robocalls. We are only aware of one other dataset of audio robocalls~\cite{robocallDatasetTechReport}, which contains robocall audio recordings from the US published by the FTC.
\end{itemize}

%% file: text/background.tex
\section{Background}

In this section, we provide more information about robocalls themselves: how they have become such a big issue and what consequences they have caused. We also introduce the main terminology used and then follow up with latest reports and statistics on robocall activity, providing an up-to-date overview of the current state of robocalls. The latter part is somewhat less formal and will include references to less-scientific resources, but still reliable sources such as FBI reports, national television reports, police station websites, etc. We still consider such information valuable given the impact robocalls have on society and given that such sources are the primary source of information for the general public.

In Section~\ref{section:related-work}, we provide a comprehensive review of the existing scientific work on robocalls. As may be observed, referenced papers also heavily rely on less-scientific resources, which is to no surprise given the practical aspect of robocalls and the heavy involvement of (non)-government institutions and organizations to help combat the problem.

\subsection{Robocall Background and Terminology}

Nassar et al. started their 2007 work with the sentence: ``\textit{Even if VoIP attacks are not in the headline news in security reviews yet, they soon will be of major harmfulness for the Internet telephony services}''~\cite{Nassar2007voiphoneypot}. VoIP is short for Voice over IP, the technology that enables propagation of calls over the internet, and unfortunately, as we can witness today, the prediction of the authors lived to be true.

The authors of~\cite{prasad2020s} provide a great overview of how the phone network operates, which we briefly summarize as well. When a caller dials a number, they are said to \textit{originate} the call, while their carrier is responsible for establishing the connection, which is poorly named as \textit{terminating} the connection. If the dialed number is not directly accessible to the carrier, the carrier routes the call by sending call signaling and media through one or more intermediate carriers. These may or may not operate in the same country. The most common signaling protocols are: Signaling System No. 7 (SS7)~\cite{itut1970ss7} for the Public Switched Telephone Network (PSTN) and Session Initiation Protocol (SIP)~\cite{rfc3261} for Voice over Internet Protocol (VoIP). Carriers connect through network gateways that can translate between signaling protocols.

Furthermore, the authors of~\cite{prasad2020s} explain the underlying mechanisms of how \textit{spoofing} is performed. Spoofing is a cyber tactic in which the perpetrator is identifying as another person or program, which in the context of robocalls means effectively falsely displaying the caller's number (caller ID) to the callee and is known as \textit{caller ID spoofing}.

An intuitive first reaction to the problem of robocalls might be: ``why not simply block the malicious user's number from making calls?''. The problem is that calls can be spoofed, making it very difficult to trace the actual number of the perpetrator. SS7~\cite{itut1970ss7} is a protocol designed in 1970, standardized since 1988, and it is still in use. In hindsight, it seems logical to implement a type of authorization, but neither SS7~\cite{itut1970ss7} nor SIP~\cite{rfc3261} have built-in authentication of the calling number. This is because historically it was assumed that only trusted carriers would terminate calls. Aside from the obvious reason of hiding their true identity, robocallers spoof numbers to make the call appear to originate from a legitimate source.

STIR (Secure Telephone Identity Revisited)~\cite{rfc7340,rfc8226,rfc8816} is a series of Request for Comments (RFC) standard documents by the Internet Engineering Task Force (IETF). It enriches the SIP protocol~\cite{rfc3261} by adding digital certificates to verify the caller's ID. Similarly, SHAKEN (Signature-based Handling of Asserted information using toKENs)~\cite{rfc8588} is a set of guidelines for the PSTN on how to deal with calls that have incorrect or missing STIR.

In 2020, the authors of~\cite{prasad2020s} were optimistically looking forward to STIR/SHAKEN as the first and only widespread cryptographic authentication mechanism in telephony. The FCC has enforced STIR/SHAKEN as of June 30, 2021 through the TRACED Act (Telephone Robocall Abuse Criminal Enforcement and Deterrence) which was passed by the US Congress in late 2019. Similarly, The Canadian Radio-television and Telecommunications Commission instructed telecommunication service providers (TSPs) to implement STIR/SHAKEN effective November 30, 2021~\cite{crtc2021compliance}.

These mechanisms have, frankly, not been effective. The authors of~\cite{prasad2020s} published new findings~\cite{prasad2023diving,prasad2025characterizing} in which they provide explanations as to why STIR/SHAKEN have failed to reduce the volume of robocalls. In essence, STIR/SHAKEN cannot completely authenticate calls originating outside a country or calls that transit a legacy phone circuit~\cite{prasad2023diving}, which robocallers have adapted to~\cite{prasad2025characterizing}.

In general, spoofing can be summarized as faking one's identity. The process we just described in which callers fake their identity is known as \textit{number spoofing} or \textit{caller ID spoofing} and should not be confused with \textit{voice spoofing}~\cite{rosello23_interspeech,tran24_interspeech,mahapatra25_interspeech,das25_interspeech}. The latter refers to malicious users applying AI technology to clone (spoof) a trusted voice (e.g. the voice of a family member) and trick the callee into believing they are communicating with somebody they trust, thereby potentially extorting valuable information, money, or obtaining system access. Finally, \textit{voice phishing} (also known as \textit{vishing}~\cite{lee2023real,ampel2026automatically}) is a social engineering attack that is carried out over the phone, which can, but does not necessarily include voice or caller ID spoofing. During a vishing attack, the perpetrator tries to trick the victim over the phone to obtain information, money, or to bypass authorized access. Vishing attacks, caller ID spoofing, and voice spoofing can all be considered robocalls, as per the definition of the FCC, provided that they were made with an autodialer or contain a prerecorded or artificial voice message.

\subsection{Broader Societal Impact}

\subsubsection{New insights}

In the Introduction section, we mentioned that robocalls are on the rise in the US: according to the report of the US PIRG Education Fund~\cite{pirg2025ringing}, robocalls are reaching a 6-year high monthly average of 2.56 billion, as shown in Figure~\ref{fig:number-of-robocalls-US}. This roughly translates to US citizens receiving 7.5 monthly robocalls per person, on average. These robocalls are not just harmlessly bothering citizens, as US citizens have reported a total of \USD{1.1} billion lost to fraud over phone calls in 2025, according to the report of the FTC~\cite{ftc2025fraudreports}. The same report lists that phone calls are the third most reported method of contact for fraud at 17\%, right behind text and email. However, the median amount of money lost to fraud committed over the phone peaks at \USD{\num{1835}}. For comparison, second on this list is text, for which the median money loss is \USD{\num{1000}}, which is almost double the difference. The 2025 statistics are not an anomaly, for example, reported median money losses in 2022 were also highest for telephony fraud, with the median reported loss over phone call scams amounting to \USD{1400}, according to the FTC~\cite{ftc2023report}.

The Federal Bureau of Investigation (FBI) founded its Internet Crime Complaint Center (IC3) in 2000. During its infancy, IC3 received roughly 2000 complaints every month~\cite{fbiic32025annualreport,fbiic32026annualreport}. From 2019 to 2024, the complaints were averaging more than 2000 per day. Losses reported to the FBI are on the rise: a total of \USD{16.6} billion in losses was reported in 2024 across several types of fraud with phishing and spoofing topping the list by number of complaints at more than 190 thousand, more than twice as much as extortion, the second on the list at over 86 thousand complaints~\cite{fbiic32025annualreport}. In 2025, a record \USD{20.9} billion in total losses was reported, with phishing and spoofing topping the list again with a record-high 191 thousand complaints.

The FTC's Do Not Call (DNC) registry, as of 2025, contains 258 million active DNC registrations. More than 2.6 million complaints were made during 2025, of which 61.2\% were robocalls~\cite{ftc2025dnc}. These numbers are on the rise: in 2024 the number of complaints was 2.09 million with 52.7\% robocalls; in 2023 a total of 2.12 million complaints, with robocalls contributing 54.6\%.

Interested readers are advised to consider the FCC's 2024 report, in which the FCC addresses the uses of artificial intelligence to protect consumers from unwanted robocalls, robotexts, and other harms, as well as enabling people with disabilities to make calls~\cite{fcc2024report}.

\subsubsection{A US-only problem?}

A lot of attention has been devoted to the US, which we attribute to the following reasons:

\begin{enumerate}
    \item Previous research revolved around the US, making it appealing to focus on the US so that comparisons can be drawn more easily against existing baselines. Previous research also influenced the availability of data.
    \item The US has a very broad number of transparent government and non-government organizations which contribute to the amount of publicly available data, making it easier to work with robocalls.
    \item As the data from our paper confirms as well, US citizens are by far the largest target of robocallers. Even though other countries are affected by robocalls, the degree of robocall quantity and severity is unparalled compared to the US. However, this does not mean that other countries are exempt from robocalls, or that the problem itself is insignificant.
\end{enumerate}

\subsubsection{Africa}

The Global Anti-Scam Alliance (GASA) 2025 report found that 68\% of adults in surveyed African countries encountered at least one scam, with \textbf{Kenya} (83\%), \textbf{South Africa} (77\%), and \textbf{Nigeria} (73\%) being the most affected countries~\cite{cfma-2025-gasa-africa-scams}. Voice calls contributed to 51\% of the scams.

\textbf{Kenya}'s Communications Authority reported 2.54 billion cyber threat incidents in Q1 2025, a 201.7\% increase from the previous quarter. The Computer Misuse and Cybercrimes (Amendment) Act of 2025 expanded the definition of phishing offences to explicitly cover fraudulent phone calls, broadening the scope beyond text-based digital communications~\cite{vellum-2025-kenya-cybercrimes-act}.

INTERPOL's Operation Red Card (2024)~\cite{interpol-2025-operation-red-card} resulted in over 300 arrests throughout Africa. In \textbf{South Africa}, 40 individuals were arrested and more than \num{1000} SIM cards were seized. According to a 2025 report from the Communication Risk Information Centre (COMRiC) an estimated \USD{300} million was linked to telecom fraud in South Africa, costing the country close to 1\% of its GDP~\cite{connectingafrica-2025-comric-fraud}.

In January 2025, 105 suspects were arrested, including 4 Chinese and 101 Nigerian citizens in Abuja, \textbf{Nigeria}. The fraudsters were targeting victims in Europe, primarily the United Kingdom~\cite{efcc-2025-arrests-105-abuja}. A month earlier, in December 2024, the Economic and Financial Crimes Commission (EFCC) had performed a raid, arresting 792 suspects for their alleged involvement in cryptocurrency investment fraud and romance scams, of which many of them international citizens.

\subsubsection{Asia-Pacific and Middle East}

According to the \textbf{Japan} National Police Agency (NPA), the total amount of money lost to unsolicited phone calls and social media reached its highest level ever in the first half of 2025~\cite{tokyoreporter-2025-npa-fraud-h1}. Fraudsters have collected close to \textyen59.7 billion (\USD{380} million) between January and June 2025, which is 2.6 times higher than in the same period the year before.
International calls accounted for approximately 70\% of fraud calls in 2024, rising to 80\% in the first half of 2025~\cite{japantimes-2025-tokyo-police-app}.
Over the whole of 2025, \textyen141.42 billion (\USD{900} million) in losses was attributed to telephony fraud, with victims typically being contacted via international phone calls and then directed to video chats on the messaging app Line~\cite{japantimes-2026-fraud-record-2025}. 

Voice phishing losses in \textbf{South Korea} hit a five-year high of \textwon433.8 billion (\USD{317} million) in 2024, up 14.1\% from the previous year, while per-case damage doubled to approximately \textwon19.9 million~\cite{sedaily-2026-voice-phishing-5yr-high}. SK Telecom, South Korea's largest carrier, reported blocking 250 million voice spam and scam calls in 2025~\cite{commsrisk-2025-skt-billion-blocked}. INTERPOL's Operation HAECHI V~\cite{interpol-2024-haechi-v} in 2024 targeted seven types of cyber-enabled fraud, including voice phishing. Korean and Chinese authorities dismantled a voice phishing syndicate responsible for \textwon\num{1511} billion (\USD{1.1} billion) in losses affecting over \num{1900} victims, resulting in 27 arrests.

\textbf{China.} The Ministry of Public Security reported intercepting 2.75 billion fraudulent phone calls and 2.28 billion suspicious SMS messages, and stopping 13.89 million people from being defrauded in 2023~\cite{chinadaily-2024-mps-crackdown}. From January to November 2024, more than \num{67000} individuals were indicted on telecom fraud charges, a 58.5\% increase year-on-year~\cite{spp-2025-prosecutors-telecom-fraud}. China adopted its own national caller identification standard (GB/T 43779-2024)~\cite{commsrisk-2024-china-caller-id-standard}, published in April 2024 with an implementation date of \DTMdate{2024-11-1}, as an alternative to the previously mentioned US-originated STIR/SHAKEN~\cite{rfc7340,rfc8226,rfc8816,rfc8588}.

The oldest television broadcaster in Southeast Asia, ABS-CBN, released a report in which they claim that the number of SMS scams in the \textbf{Philippines} has decreased in 2025, but that the number of scam calls rose by 78\% in Q3 of 2025~\cite{abs-cbn-report}.

In \textbf{Australia}, the Australian Competition and Consumer Commission (ACCC) reported that in 2023, Australians lost over \USD{3.1} billion to scams, with phone scams being a significant contributor~\cite{pietri2025telecom}.

\textbf{Singapore} and \textbf{Malaysia} signed a Memorandum of Understanding in 2024 to enhance cooperation against telecommunication scams~\cite{pietri2025telecom,indiplomacy2024sinagporemalaysia}.

The FBI, specifically the IC3 department, issued a public service announcement in May 2025 warning of fraud schemes targeting Middle Eastern students (from the \textbf{UAE}, \textbf{Saudi Arabia}, \textbf{Qatar}, and \textbf{Jordan}) studying in the United States, where scammers impersonate government or immigration officials and threaten deportation; government impersonation fraud cost Americans over \USD{405} million in 2024~\cite{fbi-2025-middle-east-students-scam}.

\textbf{UAE}'s largest telecom, e\&, has reported rolling out AI-based software to combat robocalls~\cite{andrew2025mavenir}. The \textbf{Saudi Arabia} Data and Artificial Intelligence Authority has issued a warning for consumers to reject telephone calls from people seeking their personal or bank details, warning users about vishing attempts~\cite{arabnews-2022-phone-bank-fraud-saudi}.

\subsubsection{South America}

Effective \DTMdate{2025-08-13}, all commercial calls in \textbf{Chile} must carry mandatory prefixes: \texttt{600} for solicited calls (from entities with which the consumer has an existing relationship) and \texttt{809} for unsolicited telemarketing or advertising calls~\cite{subtel-2025-prefijos-600-809,idtexpress-2025-chile-prefix-600-809}. The regulation, enforced by the Subsecretar\'{i}a de Telecomunicaciones (Subtel), enables consumers to instantly distinguish legitimate calls from potential spam before answering.

According to Hiya's 2025 report~\cite{hiyya-2025-brazil-spam-leader}, \textbf{Brazilians} receive on average 28 spam calls per month, close to 1 spam call per day, with 17\% of those calls estimated to be fraud calls.

According to Truecaller's annual report for 2025~\cite{truecaller-2025-report}, on average 6.8 calls out of 10 calls received by Brazilians outside of their phone contact list are spam, with the ratio being slightly higher in Chile at 7/10.

\textbf{Argentina}'s National Entity of Communications (ENACOM) enforces regulations that require telecom operators to implement systems for detecting and blocking spam calls and messages~\cite{pietri2025telecom}.

\subsubsection{Europe}

\begin{figure}[t]
  \centering
  \includegraphics[width=1.0\linewidth]{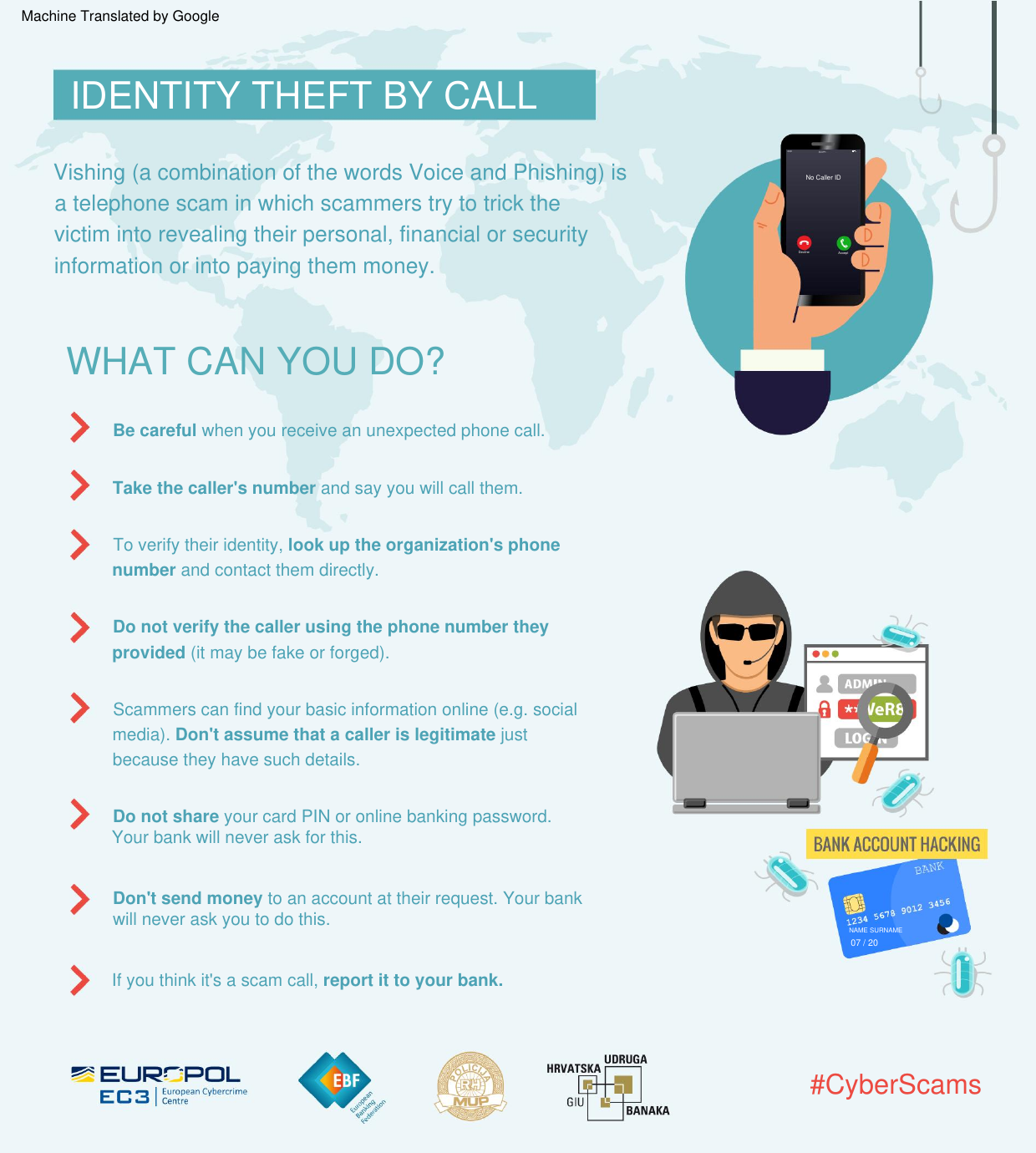}
  \caption{Info flyer published by the Croatian Police as part of Europol's campaign relating to phone theft and fraud. If the callers present themselves as an organization the police instructs citizens to not call back numbers that the caller might have specified during the call. Citizens are instead advised to look up official phone numbers of those organizations. This info flyer was automatically translated using Google, the original can be found on the Croatian Police's website~\cite{croatianpolice}.}
  \label{fig:croatian-police-vishing-flyer-translated}
\end{figure}

In \textbf{Croatia}, numerous news articles have reported incidents of fraud performed via phone robocalls. The Croatian Police~\cite{croatianpolice} published flyers instructing citizens how to behave if they receive unexpected phone calls, an example of which is shown in Figure~\ref{fig:croatian-police-vishing-flyer-translated}.

\textbf{The Netherlands.} NOS, part of the Dutch public broadcasting system, released an article in which they state that the Dutch Fraud Help Desk received nearly \num{10000} reports of scam calls that try to defraud people and extort money~\cite{nos2025scammers}. Scammers are relying on AI-enabled technology to facilitate such calls.

The news portal Telecoms reported that O2, the largest telecom in the \textbf{United Kingdom}, detected 150 million suspected scam and spam calls and blocked millions more that were later confirmed as fraudulent~\cite{nick2025o2flags}. More than 50 million suspected scam calls are flagged every month with the help of AI-based detection software, the details of which were not disclosed.

\textbf{Spain}'s largest telecommunication company Telefonica stated that they are blocking almost half a million spam or fraudulent calls daily in 2026~\cite{telefonica2026combatingfraud}.

Since \DTMdate{2025-11-19}, \textbf{Italian} telephone operators have been rejecting on average 7.5 million calls every day that are disguised as Italian numbers, but actually originate from abroad ~\cite{ilsole24ore2025harrasing}.

Obviously, we cannot fit findings for each country in this paper, but given the reports on this small sample of countries, we conclude that robocalls and spam/fraud calls in general, are a worldwide problem. Throughout this paper, we will quantify the extent to which robocalls are present in countries other than the US.

%% file: text/related_work.tex
\section{Related Work}
\label{section:related-work}

Robocalls are not a recently emerging issue. Although the term \textit{robocall} might be relatively novel, unsolicited and malicious phone calls have been well-known for some time. Researchers from North Carolina State University (NCSU) provided a significant contribution and effectively laid the foundational work on the topic of robocalls. Over the years, they published several papers on the topic of robocalls, and more generally on voice and SMS security~\cite{prasad2020s,prasad2023diving,prasad2025characterizing,reaves2016sending,reaves2016authloop,reaves2017authenticall,nahapetyan2024sms,adei2024jager}. A limitation of their work is that it considers US data only.

The 2016 systematization of knowledge paper by Tu et al.~\cite{tu2016sok} provides an overview of robocalls up to that point in time. The number of complaints, which we previously referred to be 253 million~\cite{ftc2025dnc} was 22 million at their time of writing. The paper focuses predominantly on the US.

A recent 2025 paper by Pietri et al.~\cite{pietri2025telecom} provides a more up-to-date overview of telecom spam and scam, from a perspective of the advances in AI. The paper provides a perspective on several countries across all continents. As an example, the authors show that South Africa, Nigeria, Kenya, and Ghana, have made advances such as: passing bills that mandate TSPs (Telecommunications Service Providers) to verify user identities, requiring that telecom operators implement filters/systems to block spam calls, or enforcing that TSPs maintain a registry for consumers to opt out of unsolicited communications.

\subsection{Honeypot methods}

Honeypots, in general, are systems designed to record undesirable user behavior or, in certain cases, even to attract such behavior, hence the name \textit{honeypot}~\cite{provos2004virtual}. They have no production value and are based on the assumption that any interaction with them is unauthorized, which means that the traffic that they receive is usually a probe, scan, or attack~\cite{spitzner2003honeynet}. A \textit{honeypot} may have different definitions depending on the use case, but in the context of robocalls, it can be considered as a collection of phone numbers that deliberately answer incoming phone calls from unknown callers, as these are more likely to originate from robocallers. In the late 2000s and early 2010s~\cite{gupta2015phoneypot,nassar2007,jiang2013greystar,keromytis514,carmo2011,mulliner2011poster,nawrocki2016survey}, designing VoIP honeypots was a popular research topic. 

In their 2004 paper, Pouget and Dacier~\cite{pouget2004honeypot} used a honeypot to, among other things, reduce spam activity and deceive hackers. Honeypots were utilized by Google~\cite{provos2004virtual}, researchers from Deutsche Telekom and TU Berlin~\cite{mulliner2011poster}, and researchers from the NCSU~\cite{prasad2020s,prasad2023diving,prasad2025characterizing}.

In~\cite{gupta2015phoneypot}, the authors monitored \num{39696} phone numbers. They received a total of 1.3 million calls from 250 thousand unique sources over a period of seven weeks. They detected several debt collectors and telemarketer calling patterns and one instance of a telephony denial-of-service attack. 

We already referenced several aspects of the 2020 paper from researchers of the NCSU~\cite{prasad2020s}, given that the authors provide explanations of relevant concepts such as robocalls, spoofing, STIR/SHAKEN, and similar. This paper was at the time of writing the first paper in which a large-scale honeypot was employed. The authors recorded unsolicited calls to a honeypot of up to \num{66606} lines over 11 months. The paper features a detailed analysis of the Call Detail Records (CDRs) for all calls and call content. CDRs contain a brief summary of the call, such as callee and caller number (which could be spoofed), completion status, start and end timestamps, etc. During the 11-month recording period, the authors collected \num{1481201} unsolicited phone calls. They provide further details on how they processed a large number of audio files by leveraging audio processing techniques, mainly silence detection (voice activity detection) and audio fingerprinting.

In their 2023 paper~\cite{prasad2023diving}, researchers from NCSU collected \num{232723} robocalls over a 23-month period. They present a framework called \textit{SnorCall} which enables scalable and efficient extraction of content from robocalls. In a 2025 continuation~\cite{prasad2025characterizing}, the authors tackle a significantly larger set of about 3 million voice calls. The main finding of this paper is that robocalls were in decline (which was true at the time), but that robocallers had managed to adapt to STIR/SHAKEN.

\subsection{Detecting Fraud in Telephony}

In~\cite{sahin2017using}, the authors present a bot named \textit{Lenny} that answers inbound calls and can interact with callers using a set of prerecorded voice messages. It was capable of conversing for tens of minutes while only being clearly recognized as a bot in 5\% of the calls. Similarly, the authors of~\cite{pandit2021} detect mass robocalls using a virtual assistant as well. The assistant intercepts incoming calls and asks callers to identify themselves before forwarding the call to the callee, achieving an accuracy of about 93\%. \textit{RoboHalt}~\cite{pandit2023combating} is a continuation of~\cite{pandit2021}; the authors design a virtual and interactive assistant named \textit{RoboHalt} that answers inbound phone calls on behalf of the callee and holds a natural conversation with the caller. It relies on multiple NLP (natural language processing) machine learning models to detect whether the caller is a human or a robocaller and works regardless of whether the caller is employing caller ID or voice spoofing. The system achieves a combined accuracy of 88.7\% on various types of robocalls.

Telesonar~\cite{yan2022telesonar} is an AI-powered robocall detector, that relies on echo channel detection and breath sound timing, used to determine whether speech is AI-generated or human. Elizalde and Emmanouilidou~\cite{elizalde2021detection} proposed a method for discerning between computer-generated and human-generated voices, as well as between spam and non-spam audio. They compare methods that directly operate on spectro-temporal features against methods that operate on features extracted using the OpenSmile library~\cite{eyben2010opensmile}. In both of these approaches, robocalls are only considered to be AI-synthesized speech, while prerecorded messages do not count as robocalls.

In~\cite{natarajan2021spam}, the authors transcribe calls and convert them into a high-dimensional latent space, after which they classify them as either spam or ham. Lee and Park~\cite{lee2023real} proposed a model for real-time vishing detection in Korean, which is also based on the transcript of the call.

In~\cite{kapourniotis2011scam}, the authors present a fraud detection framework for VoIP calls. They use a Bayesian network to determine whether a call is fraudulent or not by only analyzing CDR data. Similarly, Jiang et al. present a method for detecting voice-related fraud activities using only call detail records~\cite{jiang2012isolating}. Their approach uses \textit{voice call graphs}, which represent voice calls from domestic callers to foreign recipients, and propose a Markov clustering-based method for isolating dominant fraud activities from international calls. They collected data over a two year period from one of the largest cellular networks in the US.

Tas and Baktir~\cite{tas2024blockchain} propose a blockchain-based caller-ID authentication method, that would prevent caller ID spoofing in real-time. A similar approach that utilizes the blockchain for identity authentication is proposed by Chen et al.~\cite{chen2021callchain}. Tseng et al. propose FrauDetector~\cite{tseng2015fraudetector}, a graph-mining-based framework for fraudulent phone detection. Although the paper does show promising results, it does not take into account spoofing attacks. 

Research on attacks facilitated through SMS is closely related to robocalls~\cite{reaves2016sending,nahapetyan2024sms,jiang2013greystar,mishra2020smishing,reaves2018characterizing,abdulhamid2017review}. What makes robocalls more challenging is~\cite{tu2016sok}:
\begin{enumerate}
    \item \textit{Immediacy.} Call requests are immediate, and therefore anti-spam systems must complete analysis within a short window of time.
    \item \textit{Audio streams.} The spoken content of a call is not known in advance, unlike in text-based communication, therefore, any content-based analysis must be performed in real-time in a streaming fashion, while detection methods must be causal.
\end{enumerate}

%% file: text/method.tex
\section{Method}
\label{section::method}

In this section, we describe the method that we used to record robocall activity. All calls were recorded using Infobip\footnote{\url{https://www.infobip.com/}}, an international communication platform. Infobip provided us with the phone numbers and hardware to record the calls, under certain ethical agreements and legal obligations that we discuss in the following subsection.

\begin{figure}[!h]
    \centering
    \begin{subfigure}[t]{0.48\textwidth}
        \centering
        \includegraphics[width=\linewidth, trim=50pt 100pt 50pt 90pt, clip]{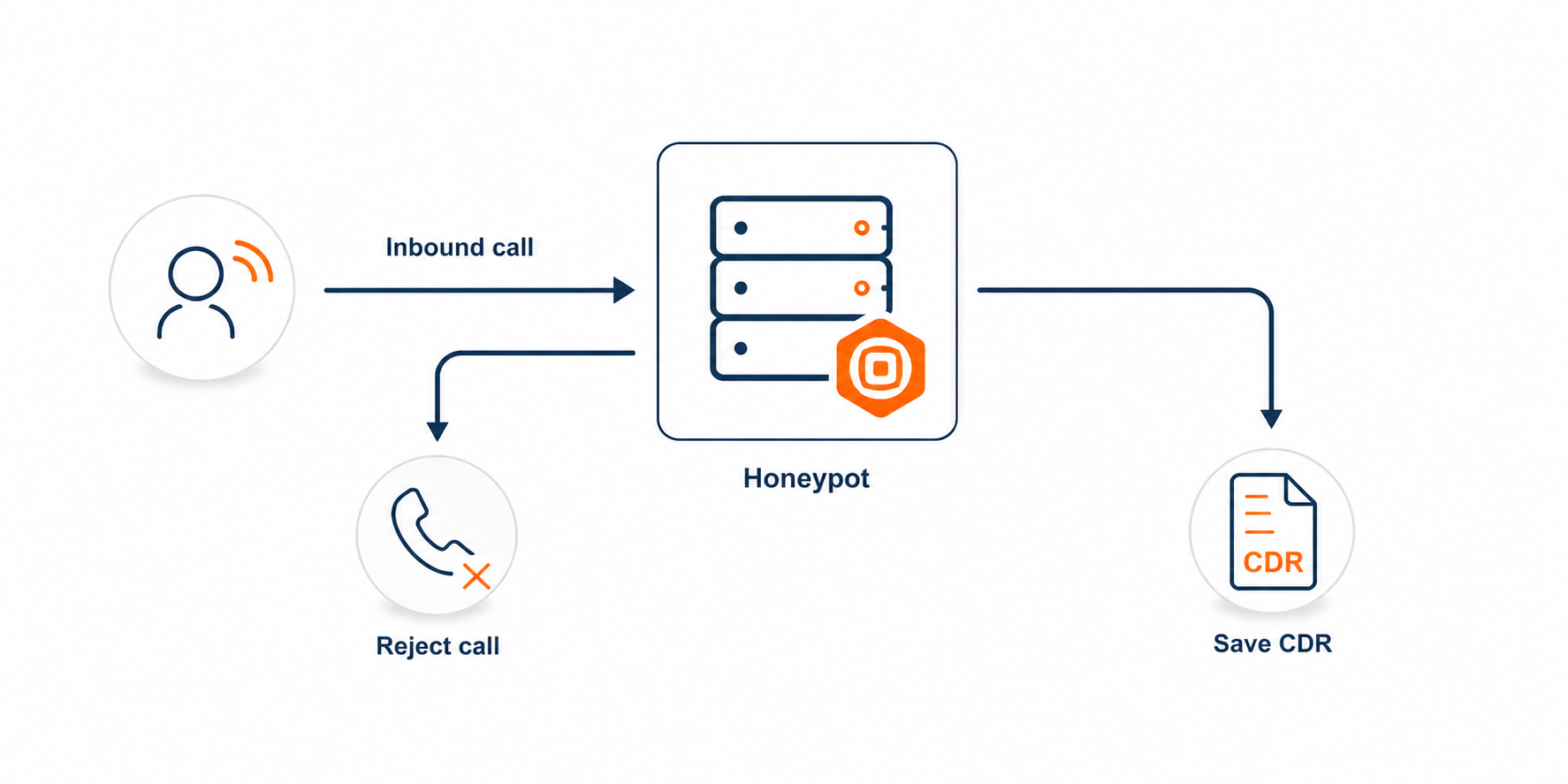}
        \caption{Honeypot method 1: passive (non-interfering mode). Incoming calls are simply rejected, but the caller ID, destination phone number, and timestamp are registered and added to the database.}
        \label{fig:honeypot-method-passive-cdr}
    \end{subfigure}
    \vfill
    \begin{subfigure}[t]{0.48\textwidth}
        \centering
        \includegraphics[width=\linewidth, trim=50pt 250pt 50pt 90pt, clip]{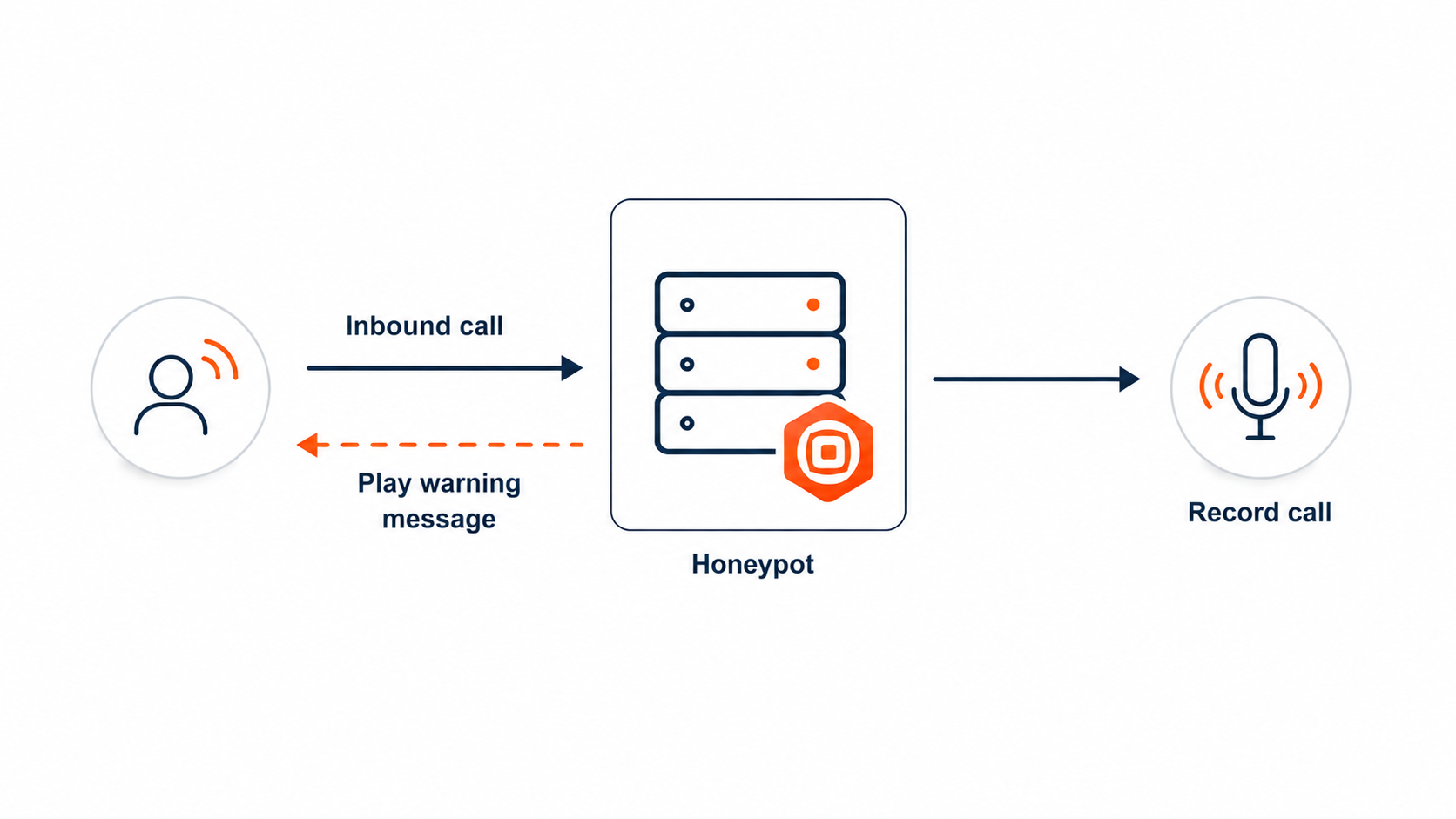}
        \caption{Honeypot method 2: interfering mode. Based on the caller's country, a warning message is played to them in their country's official language to inform the calling party that the call is being recorded. Callers are given enough time to hang up to avoid being recorded.}
        \label{fig:honeypot-method-interfering-recording}
    \end{subfigure}
    \caption{Two honeypot configurations used for collecting data.}
    \label{fig:honeypot-method-entire-fig}
\end{figure}

Figure~\ref{fig:honeypot-method-entire-fig} provides a simplified diagram of two different honeypot configurations. We used the same honeypot consisting of the same phone numbers, but in two different configurations (setups/modes).

The first setup is termed \textit{passive}, which corresponds to Figure~\ref{fig:honeypot-method-passive-cdr}. In this scenario, any incoming calls are simply rejected by the honeypot, and the honeypot does not interact with the calls. However, the honeypot registers the basic details of the call in a database. These are the previously mentioned CDRs, which contain basic call information (caller ID, callee ID, timestamp, and duration -- in this case, always 0).

The second honeypot setup is the \textit{interfering setup}, shown in Figure~\ref{fig:honeypot-method-interfering-recording}. In this scenario, the honeypot answers the incoming call and plays a warning message (an announcement) to the caller, based on the caller's country prefix, to inform them about the recording taking place.

\begin{figure*}[!ht]
  \centering
  \includegraphics[width=1.0\linewidth]{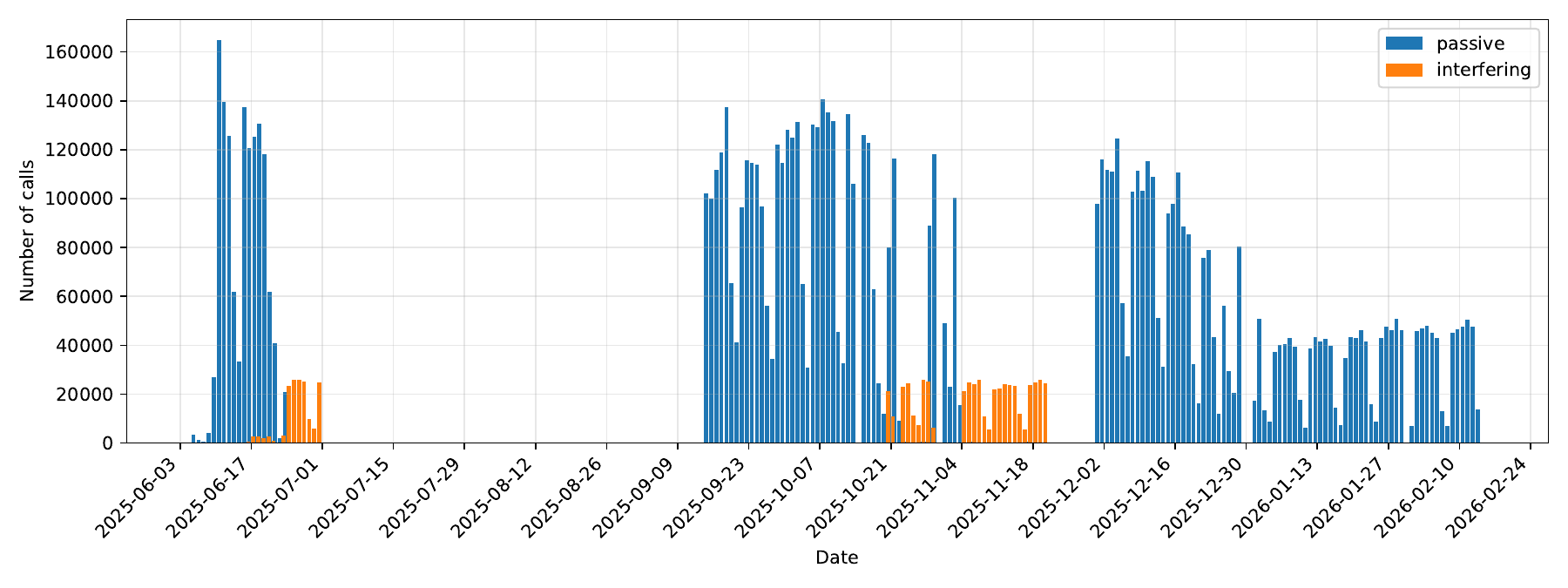}
  \caption{Daily volume of phone calls registered by the honeypot. Days colored in blue represent calls registered while the honeypot was operating in \textit{passive} mode, i.e. the honeypot does not interact with the calls, it simply rejects them. Days colored in orange indicate the \textit{interfering} mode, i.e. the honeypot answers calls and plays warning messages to inform callers about the recording process. We noticed a pattern that interacting with robocalls resulted in significantly fewer robocalls. This may imply that robocallers are analyzing the response type, which we analyze in detail later.}
  \label{fig:call-volumes}
\end{figure*}

The honeypot was active from \DTMdate{2025-06-06}, until \DTMdate{2026-02-14}, including one longer and one shorter period of inactivity. The first period of inactivity occurred during summer, i.e., \DTMdate{2025-07-01} until \DTMdate{2025-09-15}, because of a lack of personnel who could monitor the honeypot. A short 10-day period of inactivity occurred during late November, a well-known holiday period (Thanksgiving, Black Friday, Cyber Monday), as we could not guarantee the stability of the honeypot during that period given the expected increase in call volume. The number of calls received by the honeypot per day is shown on Figure~\ref{fig:call-volumes}. Calls registered while the honeypot was running in passive mode are marked in blue, while calls registered while the honeypot was running in interfering mode are marked in orange.

\subsection{Legal and ethical considerations}

Similar to previous experiments involving honeypots, we had to consider and comply with several ethical and legal requirements~\cite{prasad2020s,prasad2023diving,prasad2025characterizing}. From an ethical standpoint, our ultimate objective is to advance the knowledge of robocalls, and hopefully, contribute to a more reliable and secure telephone network. All content that we release will be publicly available for non-commercial use. We consulted with our legal department and several ethics boards in Austria and Croatia to ensure that our work is ethical and within valid legal frameworks. We have not advertised our robocall honeypot, nor did we make outbound calls. Ideally, this would mean that we should not receive any calls to our honeypot except from robocallers running mass campaigns. However, it can be the case that an individual either calls one of the available numbers in the honeypot or that a number has changed ownership in the meantime, thus we cannot assume that all calls the honeypot receives are robocalls, even though a large majority will be.

\subsubsection{CDR anonymity}
When operating the honeypot in its passive mode, we only register the caller's ID and country (and state, for US calls). Furthermore, we assign a unique ID to each caller ID and perform all downstream tasks on those unique IDs. This means that we can still analyze and differentiate between callers, but do not know their true identity (instead of operating on actual phone numbers, we operate on unique IDs). This is a safety mechanism to ensure that even if our honeypot receives legitimate calls from live humans, their caller ID will not be revealed. Furthermore, after consulting with our legal and ethics staff, we agreed to delete all call records for which the unique ID occurred only once, as these are more likely to have originated from live humans who potentially misdialed a number from our honeypot. For simplicity, we will heareafter refer to the unique IDs as the caller ID.

\subsubsection{Recording}

The interfering mode of our honeypot is more delicate from an ethical and legal standpoint, as it requires recording calls. For a majority of countries, it is sufficient for one party of a phone call to consent to recording for the recording process to be legal. However, to minimize any potential risks, we still explicitly inform the calling party about the recording process and provide them sufficient time to hang up, before the recording takes place. We play a warning message in the language of the caller's country to warn them that the call will be recorded. If the caller does not hang up within 5 seconds, the call is recorded. The template message played to the calling parties is the following:

    \begin{displayquote}
        ``You have dialed a number that is not in service. This call is being recorded to prevent spam and robocall scams over the telephone network, and the recording may be shared with partners for this purpose. If you do not want this call to be recorded, hang up within the next 5 seconds.''
    \end{displayquote}

\input{graphics/tables/phone_number_variability}

Despite the warning message, we still introduce four additional safety mechanisms:

\begin{enumerate}
    \item No metadata about recordings will be published. The recordings are anonymized and assigned a unique, randomly generated ID. They cannot be traced back to any caller ID or caller country.
    \item Any published recording must be verified by a human. The number of robocall recordings that we make publicly available (677) is significantly smaller than the number of CDRs (more than 8.7 million). If a recording is included in our dataset, it is confirmed by at least one human evaluator to be a robocall prior to publishing.
    \item All other (unverified) recordings were filtered in a process we describe later to minimize the chance that they are accidental calls made by humans. They were used for automated computer analysis and deleted afterwards, ensuring that they can no longer be accessed and were only used for the purposes of analysis in this paper.
    \item Finally, any recordings shorter than 5 seconds were automatically deleted to further ensure that accidental misdials are excluded in our research.
\end{enumerate}

Overall, we approached the sensitive topic of collecting robocall data with extreme caution and have taken several steps to ensure no violation of caller privacy and to minimize the chance of accidental/live calls made by humans being caught by our honeypot.

\subsection{Transcripts}

Similar to the recordings, we release a collection of 839 call transcripts. These were verified by at least one human annotator to be robocalls and have been processed to remove any personal identifiers. Call transcripts are released without accompanying metadata, so they cannot be traced back to the caller. Transcripts are released as a part of our automated robocall clustering and campaign analysis, which we cover in the next sections.

\subsection{Limitations}

Our honeypot has three main limitations:

\begin{itemize}
    \item \textbf{Warning messages.} Playing the warning message may influence robocallers, e.g., robocallers might be reluctant to call the same or similar numbers once aware of the recording. We consider this effect in more detail later on.
    \item \textbf{Periods of inactivity.} Periods of inactivity interrupt the continuity of the honeypot. It would be better if our data were continuously collected throughout the entire recording process.
    \item \textbf{Variability of available phone numbers.} A technical limitation of our study was that the quantity of available phone numbers varied slightly per day. This variation is relatively small, on average 2.66\% for US telephone numbers and 0.05\% for all other phone numbers. More details are provided in Table~\ref{tab:phone-number-variability}.
\end{itemize}

%% file: graphics/tables/phone_number_variability.tex
\begin{table*}
    \centering
    \caption{Quantity of phone numbers employed in the honeypot. Although the total amount of available phone numbers varied per day, this variance manifested steadily and on average resulted in a change of 2.27\% for US phone numbers and 0.05\% for other phone numbers.}
    \resizebox{0.75\linewidth}{!}{%
\centering
\begin{tabular}{l|r|r|r|r|r|r}
\hline
\multicolumn{1}{c|}{Region} & \multicolumn{1}{c|}{Daily median} & \multicolumn{1}{c|}{Min.} & \multicolumn{1}{c|}{Max.} & \multicolumn{1}{c|}{Q1} & \multicolumn{1}{c|}{Q3} & \multicolumn{1}{c}{Mean daily change}  \\ 
\hline
US                          & \num{11751}                                          & \num{9190}                      & \num{15344}                     & \num{10812}                   & \num{13243}                   & 266.54 (2.27\%)                        \\
International               & \num{101913}                                         & \num{100122}                   & \num{102375}                   & \num{101711}                 & \num{102055}                 & 54.66 (0.05\%)                          
\end{tabular}
    }
    \label{tab:phone-number-variability}
\end{table*}

%% file: text/metadata_analysis.tex
\section{Call Metadata Analysis}
\label{sec:analysis}

This section presents the analysis of robocall metadata collected from our honeypot infrastructure.
A key consideration to keep in mind is caller ID spoofing.
Robocallers routinely falsify the displayed phone number, so a call may appear to originate from the callee's country, to increase the level of trust.
Because caller metadata is unreliable, we anchor our temporal and geographic analyses on callee location, the known locations of our honeypot numbers, framing results in terms of when and where robocallers target victims rather than where they operate from.

\subsection{Call Patterns}
\label{subsection:call-patterns}

\subsubsection{Dataset Overview}

The honeypot network received \num{9656257} raw CDRs between June 2025 and February 2026.
We applied two preprocessing filters to isolate robocall activity: (1) we selected calls with duration equal to zero (passive/rejected calls), as answered calls were processed separately for audio analysis; and (2) we retained calls from phone numbers that appeared at least twice in the dataset, filtering out likely accidental human dials.

After filtering, the analysis dataset contains \num{8747351} calls from \num{573823} unique caller IDs to \num{249819} unique callee numbers.
On average, this amounts to \num{34438} calls per day across the honeypot network during the 254-day collection period.

In the filtered metadata dataset (duration-zero calls from caller IDs observed at least twice), \num{8314813} calls targeted US callee numbers (95.1\%), while \num{432538} calls targeted non-US callee numbers (4.9\%). These calls reached \num{240163} distinct US callee identifiers and \num{9656} distinct non-US callee identifiers, corresponding to 34.6 and 44.8 calls per reached callee, respectively.
This normalization counts only honeypot numbers that received at least one retained call. It is not the total supervised inventory and does not refer to calls that received the warning message. Availability snapshots show approximately \num{12006} US numbers and \num{101738} non-US numbers on average. Normalized by this inventory, US numbers received substantially more traffic: 692.6 calls per available US number versus 4.25 per available non-US number over the collection period. Thus, the higher international value reflects concentration among reached international numbers, not a higher overall targeting rate.

%US-based honeypot numbers received 8,314,813 calls (95.1\%), while international numbers received 432,538 calls (4.9\%).
%The US portion of the honeypot that had received calls contained 240,163 numbers compared to 9,656 international numbers.
%On a per-number basis, US honeypots received 34.6 calls per number over the collection period, while international honeypots received 44.8 calls per number.
%The higher per-number rate for international numbers is notable given that US robocall activity is often considered uniquely severe. 

\subsubsection{Temporal Patterns}

\begin{figure}[ht]
\centering
\includegraphics[width=\linewidth]{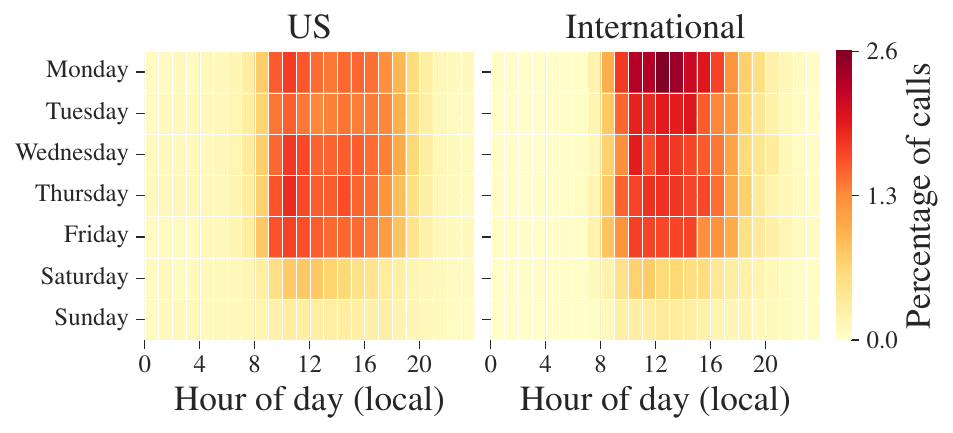}
\caption{Temporal distribution of robocalls by hour of day and day of week for US (left) and international (right) callees. Timestamps are in callee local time. Color intensity indicates the percentage of total calls in each region.}
\label{fig:temporal_heatmap}
\end{figure}

Figure~\ref{fig:temporal_heatmap} shows the temporal distribution of calls by hour of day and day of week, segmented by callee region.
All timestamps were converted to the callee's local timezone to capture when robocallers target numbers in each region.

For US callees, robocall activity concentrates during business hours.
Calls to US numbers peak at 10:00 AM local time (9.7\% of daily volume), with secondary peaks at 11:00 AM and 2:00 PM.
Overall, 61.0\% of US-targeted calls occur during standard business hours (9 AM to 5 PM, Monday through Friday).
Weekend activity drops substantially, accounting for only 12.4\% of weekly volume.
Nighttime hours (11 PM to 6 AM) see minimal activity at 5.6\% of calls.
This pattern suggests automated dialing systems configured to maximize answer rates by targeting recipients during waking hours.

International callees exhibit a tighter business-hours concentration.
A higher proportion of calls (69.9\%) occur during weekday business hours, with peak activity at noon local time (10.8\% of daily volume).
Weekend calls account for only 10.0\% of volume, and nighttime calls drop to 1.8\%.
The sharper weekday concentration for international numbers may reflect more targeted campaigns or operators with better knowledge of local business customs in specific countries.

\subsubsection{Caller Concentration}

\begin{figure}[t]
\centering
\includegraphics[width=\linewidth]{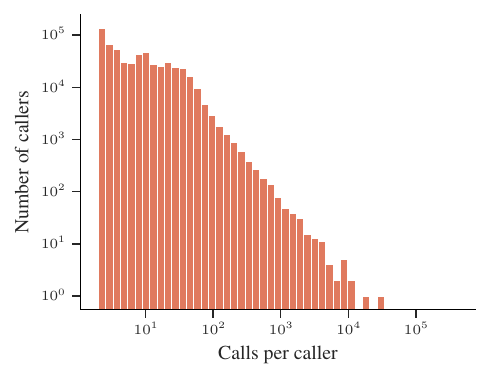}
\caption{Distribution of calls per caller ID across the full dataset (US and international data combined) on log-log scale. The heavy-tailed pattern shows that most callers make few calls, while a small number of prolific callers generate substantial traffic.}
\label{fig:caller_distribution}
\end{figure}

We analyzed the distribution of calls across all caller IDs in the dataset to understand whether robocall activity is concentrated among a few prolific actors or distributed broadly.
Figure~\ref{fig:caller_distribution} shows the distribution of calls per caller on a log-log scale.

The distribution is heavily skewed.
The median caller made only 6 calls over the 254-day collection period, while the mean was 15.2 calls, and the most prolific caller made \num{416015} calls.
Half of all callers (49.9\%) made between 2 and 5 calls, while only 0.03\% (181 callers) made more than \num{1000} calls each.
With a Gini coefficient of 0.66, where 0 represents perfect equality and 1 represents all calls originating from a single caller, it is evident that robocall activity is highly concentrated among a minority of callers.

Concentration metrics reveal that a small fraction of callers generate a disproportionate share of traffic.
The top 1\% of callers (\num{5738} caller IDs) account for 24.4\% of all calls.
The top 5\% account for 41.1\%, and the top 10\% account for 53.9\%.
This concentration suggests that targeting high-volume callers could substantially reduce robocall traffic.

Comparing US-targeting and international-targeting callers reveals structural differences in campaign organization.
US-targeting calls originate from \num{501960} unique caller IDs with a Gini coefficient of 0.66, while international-targeting calls originate from only \num{71982} caller IDs with a lower Gini of 0.53.
The lower concentration for international campaigns suggests either more distributed operations or less sophisticated automation.
Interestingly, only 119 caller IDs (0.02\%) appear in both US and international call records, indicating that robocall campaigns targeting different regions operate as almost entirely separate operations with distinct infrastructure.

\subsubsection{Geographic Coverage}

The dataset spans 178 unique caller country codes and 65 callee countries.
The callee distribution reflects our honeypot deployment: US numbers received \num{8314813} calls (95.1\%), followed by Great Britain with \num{217787} calls (2.5\%), Poland with \num{58356} calls (0.7\%), Canada with \num{33373} calls (0.4\%), and Denmark with \num{31505} calls (0.4\%).

Caller country codes are subject to spoofing and should be interpreted with caution.
In the raw data, 85.7\% of calls displayed US caller IDs, followed by Nigeria (3.9\%), Great Britain (2.4\%), Saudi Arabia (0.8\%), and Canada (0.7\%).
Examining caller-callee country pairs reveals strong evidence of neighbor spoofing: 89.9\% of all calls display a caller country matching the callee country.
The dominant pair is US to US (85.5\% of all calls), followed by Great Britain to Great Britain (2.4\%) and Poland to Poland (0.6\%).
This pattern suggests that robocallers systematically spoof local numbers to increase answer rates.

Cross-border caller IDs appear primarily for US-targeted calls.
Nigeria accounts for 3.9\% of displayed caller origins, followed by Saudi Arabia (0.7\%), Kenya (0.6\%), and Afghanistan (0.5\%).
These may represent either true call origins or deliberate foreign number display for specific scam types.
The presence of \texttt{UNKNOWN} caller IDs (0.5\%) indicates calls where caller ID information was unavailable or malformed.

\subsection{Regional Analysis}
\label{subsec:regional_analysis}

% \purple{TBA by Andro and/or Tomo. I can provide additional analysis if needed.}

We have already noted that domestic traffic makes up the largest share of the dataset, especially for the three leading same-country pairs: the US, the United Kingdom, and Poland.
Same-country calls account for \num{7866600} of \num{8747351} total calls (89.9\%), with the domestic origin being the single most frequent caller for 54 of 65 destination countries.
Given this strong domestic pattern, the next natural step is to examine the most frequent non-domestic callers in the dataset.

\begin{figure*}[t!]
\centering
\includegraphics[width=\linewidth]{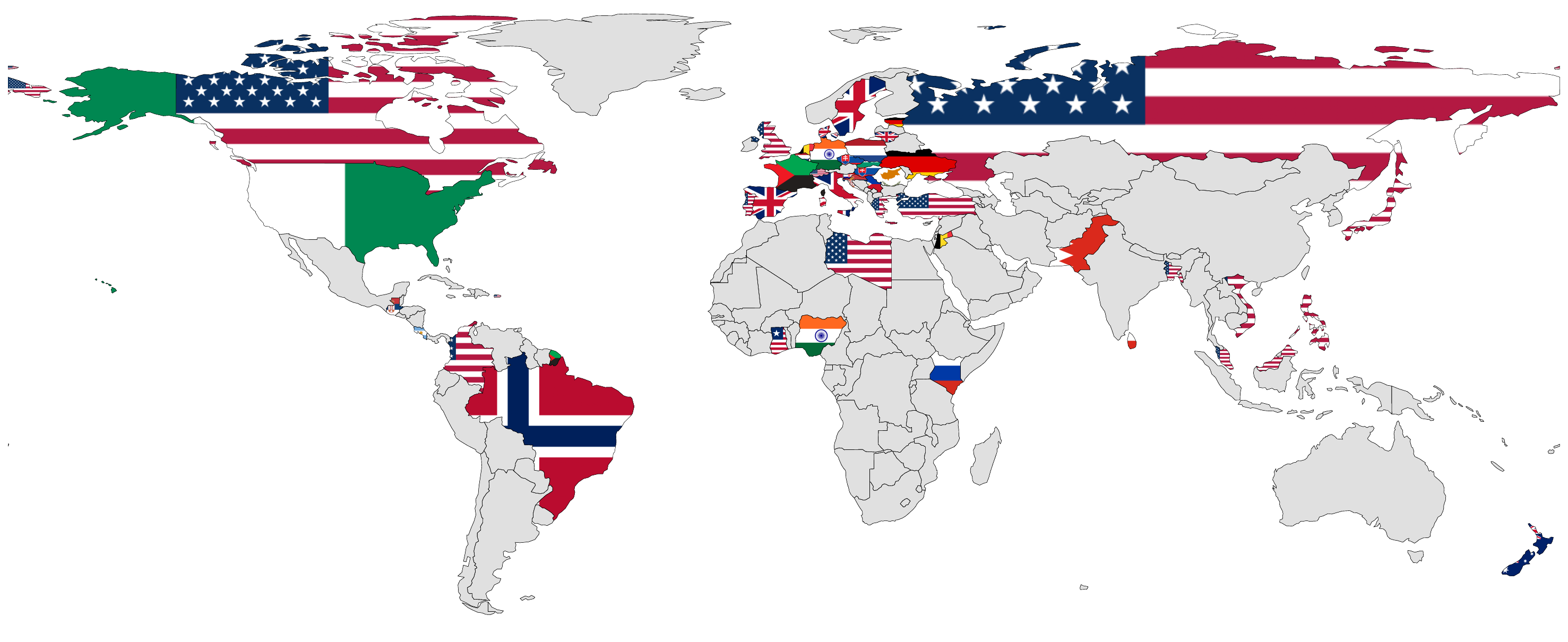}
\caption{Most frequent non-domestic caller country by destination country. The United States is the dominant external source for 16 destinations, followed by the United Kingdom for 5 destinations.}
\label{fig:world_map_top_callers}
\end{figure*}

\Cref{fig:world_map_top_callers} shows the most frequent non-domestic caller country observed in the CDRs for each destination country. 
After excluding domestic calls, a dominant external caller can still be identified for 47 destination countries, and we are left with 22 unique source (caller) countries.
The United States is the most common dominant external source, leading in 16 countries, followed by the United Kingdom in 5 destinations.
A small number of other countries, including the Netherlands, India, Slovakia, Germany, Serbia, and Belgium, are the top external caller for two destinations each, while the remaining source countries appear only once.

Here we also see the pronounced volume of calls originating from Nigeria, which is the leading external source for calls to the United States (\num{340808} of \num{787594} cross-border calls).
The United States dominates calls to Canada (\num{10306} of \num{10767}) and the Philippines (\num{576} of \num{1323}), the United Kingdom is the dominant source for Denmark (\num{3553} of \num{13253}) and Sweden (\num{47} of \num{126}), and India is the top caller for Germany (\num{72} of \num{73}).
Overall, the figure shows that international call traffic is not evenly distributed across countries, but is instead concentrated around a relatively small set of recurring country-to-country connections.

We can repeat a similar analysis for traffic to US states.
Domestic traffic is again overwhelmingly dominant, as calls from within the United States account for \num{7375854} of \num{8143011} calls to US states with a resolvable caller country (90.6\%), and the domestic source is the largest contributor in all \num{51} covered jurisdictions (50 states and the District of Columbia).
For this reason, it again makes sense to focus the analysis on non-domestic calls.

\begin{figure*}[ht]
\centering
\includegraphics[width=\linewidth]{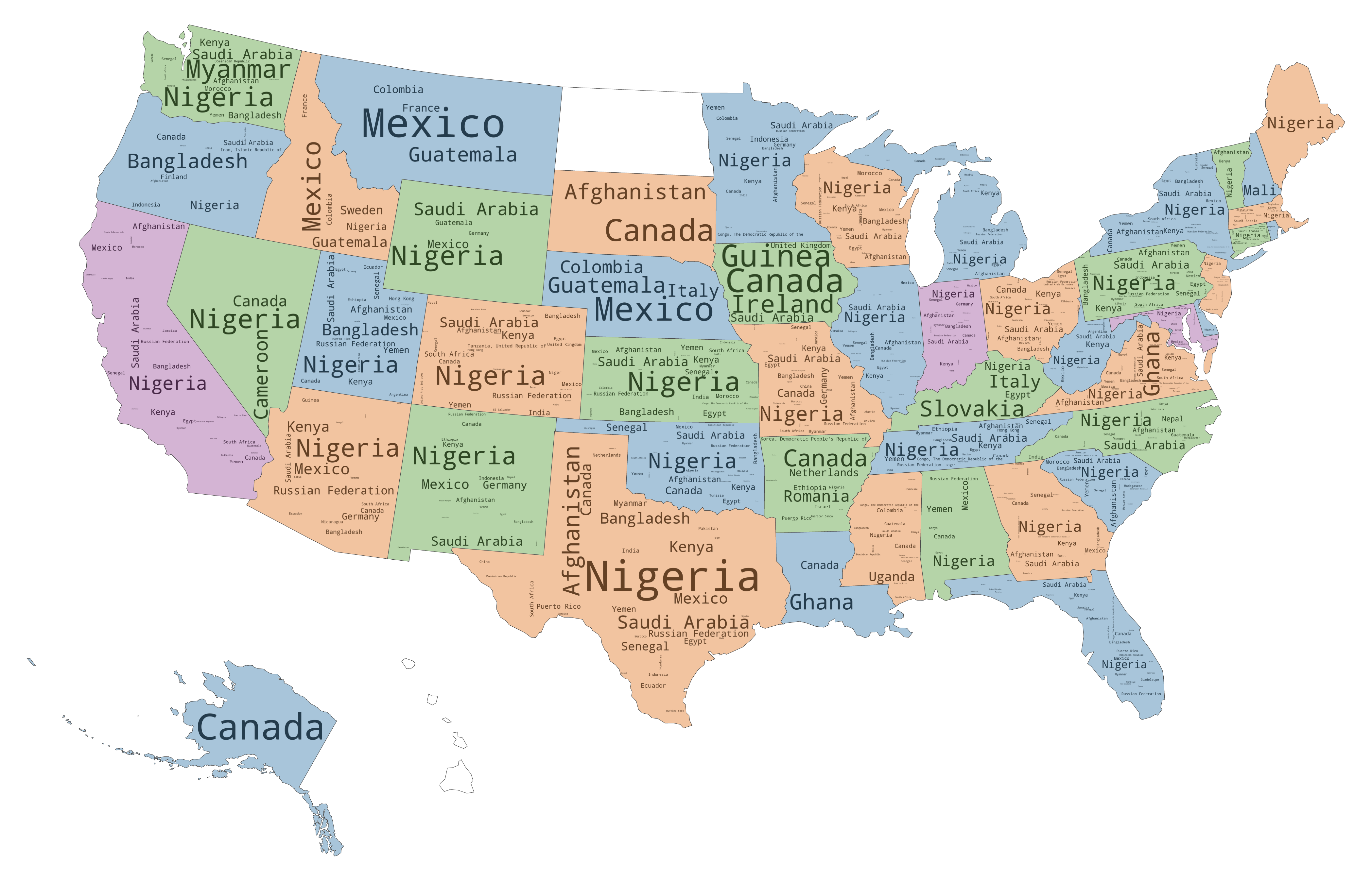}
\caption{Word clouds of caller countries by US state, with word size proportional to the number of non-domestic calls. Nigeria is the dominant external source in 35 states, followed by Canada and Mexico, each leading in 4 states.}
\label{fig:us_state_wordcloud}
\end{figure*}

\Cref{fig:us_state_wordcloud} shows the distribution of non-domestic caller countries in the CDRs for each US state, with word size indicating frequency and the largest word corresponding to the dominant external source.
Even after excluding domestic calls, the remaining external traffic is still fairly concentrated.
California has the largest absolute number of non-domestic calls (\num{123197}, 16.1\% of non-domestic calls), followed by New York (\num{66254}, 8.6\%) and Massachusetts (\num{66212}, 8.6\%), while Nigeria is the dominant external source in the largest number of states, leading in \num{35}.
Canada and Mexico each lead in \num{4} states, Ghana leads in \num{2}, and Bangladesh, Mali, Slovakia, and Uganda each lead in one state.
Note that there were no non-domestic calls recorded for North Dakota and Hawaii.
Overall, the figure shows that non-domestic call traffic to the United States is concentrated among a relatively small set of external origins rather than being evenly distributed across states.

\subsection{Temporal Co-targeting and Transcript Similarity}
 \label{subsec:temporal-cotargeting}
 
Displayed caller IDs are observable in the CDRs, but they are not reliable actor identifiers, because robocallers can spoof them. We therefore tested whether call timing and shared callees reveal structure that is also visible in call content. We proceed in three steps. First, we link caller IDs that target the same numbers at around the same time. Second, we group those links into communities. Third, we test whether callers in the same group also deliver similar spoken messages. The first two steps use only call metadata; the third brings in transcript content as an independent check, since metadata and message content are separate signals.

For this analysis, we reused the CDR preprocessing from the previous metadata analysis but retained answered calls, so that metadata groups could be compared with transcripts. Regions were assigned by callee country. Before constructing the graph, we removed the ten most widely targeted callee numbers per region, measured by the number of distinct caller IDs that called each callee. Otherwise, these callees would connect many unrelated callers.

\begin{figure}[t]
\centering
\includegraphics[width=\linewidth]{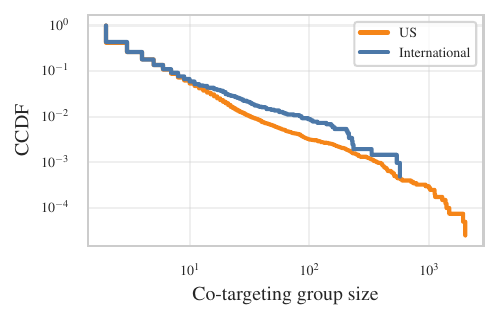}
\caption{Complementary CDF of temporal co-targeting group sizes for US and international callees. Both distributions are heavy-tailed: a large majority of groups contain only a few caller IDs, while a small number of groups exceed $10^2$ members. Groups of size 1 are excluded.}
\label{fig:temporal-cotargeting-group-sizes}
\end{figure}

The intuition behind the first step is that caller IDs belonging to one campaign tend to dial the same honeypot numbers at around the same time. We therefore built an undirected co-targeting graph separately for US and international callees, in which each node is a caller ID. We connected two caller IDs when they called the same honeypot number within a 24-hour window. We weighted each edge by the number of distinct callees the two shared, so that callers linked through many common targets count as more strongly related. We then grouped the connected caller IDs using weighted Leiden community detection~\cite{traag2019louvain}, a standard approach that partitions a graph into densely connected communities. The resulting US graph consisted of \num{201913} nodes and \num{702984} edges; the international graph consisted of \num{12255} nodes and \num{33812} edges. The detected communities matched connected components almost exactly, at 99.7\% in the US graph and 99.1\% in the international graph. In other words, callers separate into distinct clusters on timing and shared targets alone, before any content is considered, and Leiden only contributes meaningful splits inside a small number of larger components. The resulting group-size distribution (Figure~\ref{fig:temporal-cotargeting-group-sizes}) is heavy-tailed in both regions: most groups contain only several caller IDs, while a small number of groups are large.

Having grouped callers by metadata alone, we next asked whether those same groups are reflected in what the callers actually say. Each transcript was embedded with BGE-M3~\cite{bge-m3}, a multilingual embedding model, that maps each transcript to a vector so that texts with similar meaning lie close together. We use a multilingual model, because the transcript set spans multiple languages and we compare US and international calls in the same analysis. For caller IDs with multiple transcripts, we averaged the transcript embeddings to obtain one caller-level vector. This avoids comparing repeated transcripts from the same caller, so that all within-group comparisons are between distinct caller IDs. The validation subset comprises graph nodes with at least one transcript passing our cleaning filter (length above 20 characters) that also fall in a group with at least one other transcript-covered caller. This left \num{1459} US caller IDs across 217 groups and 123 international caller IDs across 21 groups.

The test is straightforward. If co-targeting groups correspond to coordinated campaigns, callers inside the same group should sound more alike than callers drawn from different groups. For each group, we computed the mean cosine similarity over all within-group caller pairs. Formally, for a group $g$ with caller-level vectors $\{\mathbf{v}_i\}$,
\begin{equation}
\bar{s}_{\text{within}}(g) = \frac{1}{\binom{|g|}{2}} \sum_{i < j} \cos(\mathbf{v}_i, \mathbf{v}_j),
\qquad
\cos(\mathbf{v}_i, \mathbf{v}_j) = \frac{\mathbf{v}_i \cdot \mathbf{v}_j}{\lVert \mathbf{v}_i \rVert \, \lVert \mathbf{v}_j \rVert}.
\label{eq:within-group-similarity}
\end{equation}
For the between-group baseline, we sampled, for each group, pairs between that group and all other groups. We drew ten between-group pairs for every within-group pair, capped at \num{5000} pairs per group. We then averaged the per-group within-means and per-group between-means across groups, so that large groups did not dominate. The uplift is their ratio,
\begin{equation}
\text{uplift} = \frac{\overline{s}_{\text{within}}}{\overline{s}_{\text{between}}},
\label{eq:uplift}
\end{equation}
where $\overline{s}_{\text{within}}$ and $\overline{s}_{\text{between}}$ denote the group-averaged within- and between-group mean similarities.
Confidence intervals were obtained by resampling groups with replacement
(\num{2000} bootstrap iterations) and recomputing the ratio of the two averages on each resample.

\begin{figure}[t]
\centering
\includegraphics[width=\linewidth]{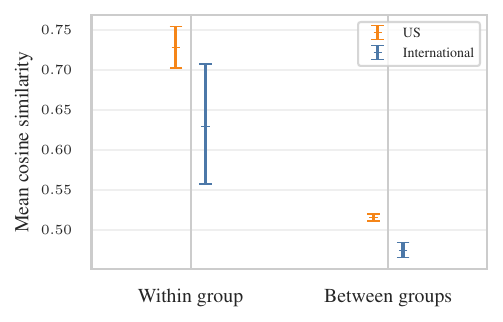}
\caption{Group-level cosine similarity of caller-averaged BGE-M3 transcript embeddings, within versus between temporal co-targeting groups, for US and international callees. Markers show the mean per-group similarity averaged across groups; error bars are 95\% bootstrap intervals over groups.}
\label{fig:temporal-cotargeting-similarity}
\end{figure}

For US callers, the mean within-group similarity was 0.728 against a between-group mean of 0.516, an uplift of $1.41\times$ with a 95\% bootstrap interval of $[1.37, 1.46]$. For international callers, the corresponding means were 0.629 and 0.475, an uplift of $1.33\times$ with an interval of $[1.18, 1.50]$. Figure~\ref{fig:temporal-cotargeting-similarity} shows the per-group means together with their bootstrap intervals. In both regions, the interval lies well above one, so the higher within-group similarity is unlikely to be a chance artifact of the grouping.

The transcript comparison supports the structure found in the caller graph. Callers grouped together purely by when and whom they called also deliver measurably more similar messages than unrelated callers. We therefore treat temporal co-targeting groups as evidence of campaign-like organization in the call traffic, while avoiding one-to-one attribution of groups to campaigns because caller IDs can be spoofed and transcript coverage is incomplete.

%% file: text/content_analysis.tex
\section{Call Content Analysis}

\begin{figure}[t]
  \centering
  \includegraphics[width=1.0\linewidth]{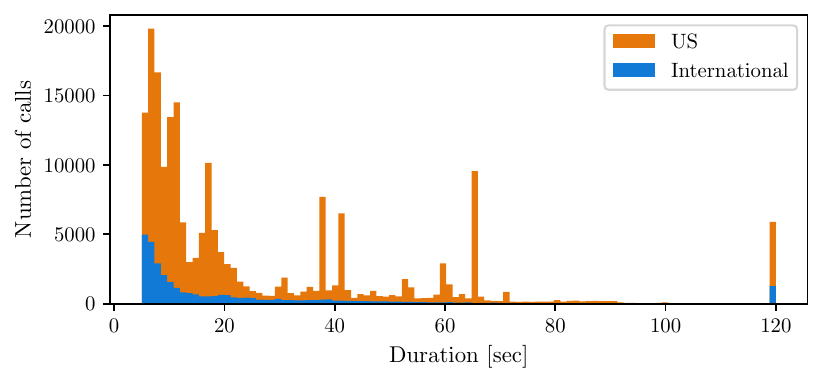}
  \caption{Duration of recorded US and international phone calls. The distributions for both regions follow a similar pattern, with most calls being shorter than 20 seconds.}
  \label{fig:call-durations}
\end{figure}

In this section, we analyze the audio content of the recorded calls. In total, we have recorded \num{220301} calls in the US, and \num{42798} calls in the rest of the world. Due to the aforementioned legal reasons, calls shorter than 5 seconds were discarded, after which \num{197765} (89.77\%) US calls were left and \num{33253} (77.7\%) international calls remained. We can therefore only speculate about the nature of the shorter calls, as we have no way of discerning what percentage of those calls belong to robocalls or to legitimate, human calls. A histogram depicting the durations of the recorded calls is shown in Figure~\ref{fig:call-durations}.

\subsection{Voice Activity Detection}

Previous work has shown that a notable percentage of unsolicited calls are silent, or contain very little spoken content. In~\cite{prasad2020s}, 62.75\% of the calls had less than 10\% spoken content, while 42.05\% had less than 1\% of audio content. In~\cite{prasad2023diving}, 31.28\% of calls over a collection of more than 1.3 million were completely silent. One possible explanation provided by~\cite{prasad2020s} was that silent calls are used to simply identify which phone numbers are active and capable of answering a phone call. Another plausible explanation could be that callers are using voice activity detection features that trigger playback of recorded messages once, and only if the caller is confident that the call has been answered by an actual person. This method is known as \textit{answering machine detection} (AMD)~\cite{gomez2014answering,altwlkany2024recurrent} and it has been used by robocallers before~\cite{tu2016sok}. This heavily aligns with our distribution of recorded calls in Figure~\ref{fig:call-volumes}. We had several switches between the passive and interfering modes of our honeypot. Once switching to the interfering mode, the volume of received calls dropped significantly, sometimes resulting in a tenfold drop. We therefore have strong reason to believe that robocallers utilize technology such as AMD to determine whether the call was picked up by a human or an answering machine. Since in our case a warning message is played (due to the legal and ethical agreements we made), robocallers could have detected this and stopped calling the given number and neighboring phone numbers.

\begin{figure}[t]
  \centering
  \includegraphics[width=1.0\linewidth]{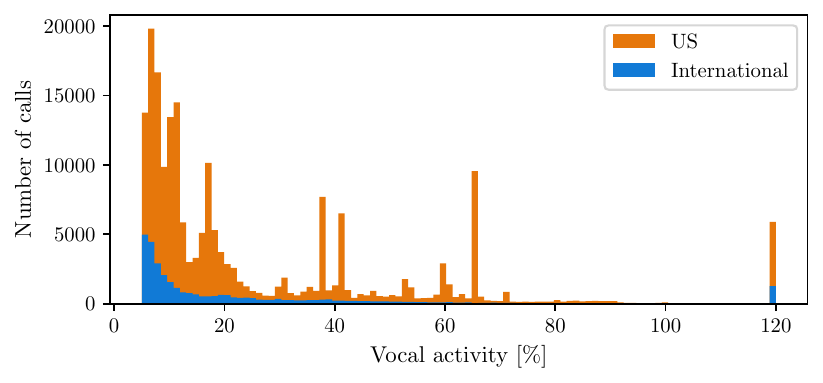}
  \caption{Percentage of vocal activity detected. In both regions, a substantial amount of robocalls are silent, with very little vocal activity. This is a common phenomenon among robocalls and has been previously observed~\cite{prasad2020s,prasad2023diving}. ``Blank'' calls are usually made with the goal of determining whether the dialed number has voice capabilities (scouting calls).}
  \label{fig:vocal-activity}
\end{figure}

\begin{figure*}[!ht]
    \centering
    \begin{subfigure}{0.48\textwidth}
        \centering
        \includegraphics[width=\linewidth]{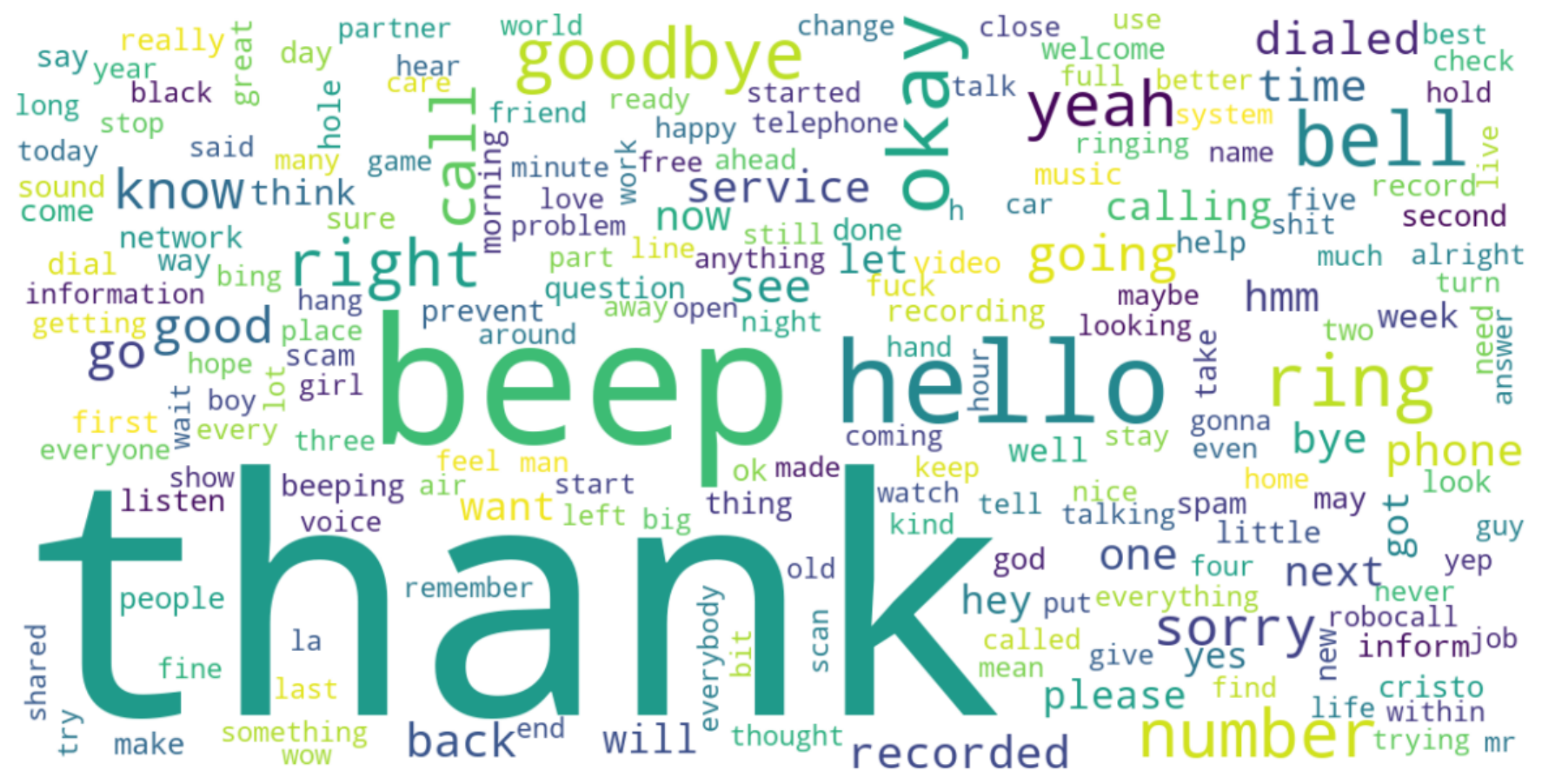}
        \caption{Wordcloud for low speech ratio recordings.}
        \label{fig:word-cloud-low-speech-ratio}
    \end{subfigure}
    \hfill
    \begin{subfigure}{0.48\textwidth}
        \centering
        \includegraphics[width=\linewidth]{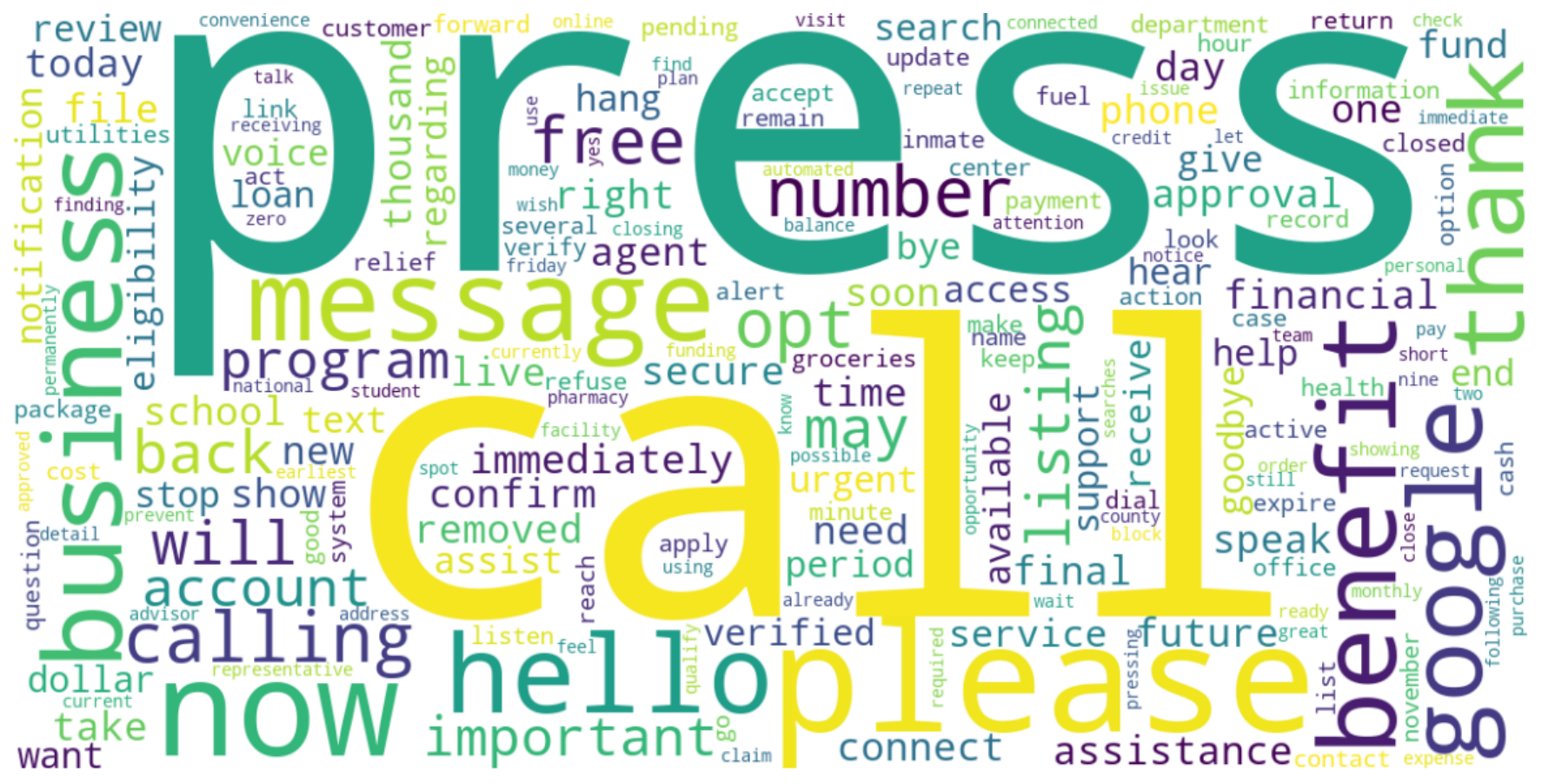}
        \caption{Worldcloud for high speech ratio recordings.}
        \label{fig:word-cloud-high-speech-ratio}
    \end{subfigure}
    \caption{Hallucinations caused by transformer-based automatic speech recognition can heavily impact the transcriptions. Hallucinations are more prominent in recordings that contain a lower ratio of speech compared to silence. To the left, we show the most common words occurring in low speech ratio recordings, while the image on the right shows the most frequently occurring words in high speech ratio recordings. There is a clear difference between the distribution of words between the two groups, and the words on the left correspond to well-known and frequent hallucinations produced by Whisper~\cite{koenecke2024careless,baranski2025investigation}.}
    \label{fig:word-clouds-low-high-speech-ratio}
\end{figure*}

We used SileroVAD~\cite{SileroVAD}, an industry-grade voice activity detection (VAD) module designed specifically for use in telephony, to detect vocal segments of audio. With SileroVAD, we observed that a total of \num{50911} (24.78\%) calls contain any vocal activity at all in the US, compared to \num{13639} (41.02\%) internationally. The relative difference between the proportions of calls with vocal activity in the US and international calls is 65.54\%, indicating that calls in the US are more likely to be silent. Given previous research findings, one explanation may be that US robocallers are simply more sophisticated and use AMD so that playback messages are only triggered once the robocaller is confident that the call was answered by an actual human. In Figure~\ref{fig:vocal-activity}, we show a histogram of the total percentage of vocal activity per recorded call.

\subsection{Transcript Analysis}

Having determined the amount of spoken content, we proceed to transcribe the calls. We transcribed all calls using Whisper~\cite{radford2023robust}, a state-of-the-art open-source automatic speech recognition model (specifically model version \textit{large-v3-turbo}). Our subsequent analysis is based on a heuristically determined threshold which is that the calls should contain at least 25\% spoken content, in line with previous work~\cite{prasad2020s}. The reason for this is that transcribing such a large number of calls is done automatically, and state-of-the-art transcription models suffer from hallucinations~\cite{koenecke2024careless,baranski2025investigation}.

We justify our decision and illustrate this by separately transcribing US recordings that contain less than 25\% of spoken content (low speech ratio) and recordings that contain more than 25\% of spoken content (high speech ratio). We extract the most commonly occurring words from both sets of recordings and represent them as word clouds in Figure~\ref{fig:word-clouds-low-high-speech-ratio}. By far the most commonly occurring word in the low speech ratio recordings is the word \textit{thank}, which among other is probably part of the most frequent hallucination produced by Whisper: \textit{"thank you"}~\cite{koenecke2024careless,baranski2025investigation}. The word \textit{beep} is the second most spoken word in the low speech ratio recordings, while it does not even appear in the word cloud of the most common words for the high speech ratio recordings. The words \textit{thank} and \textit{hello} are fairly common in both sets of recordings, but the differences are clear. Action verbs such as \textit{call} and \textit{press} top the high speech ratio recordings, followed by words that indicate a business matter or sense of urgency: \textit{important}, \textit{message}, \textit{account}, \textit{business}, \textit{google}, \textit{service}, \textit{number}, etc. We therefore limit further analysis to recordings that contain at least 25\% spoken content, leaving \num{20328} US robocalls and \num{4071} international robocalls.

\subsubsection{Most Common Languages}

In Table~\ref{tab:most-common-languages} we provide the top 5 most commonly occurring languages and the percentage of their representation within the recordings for the US and international calls. The language detection is provided by Whisper~\cite{radford2023robust}, which in addition to the transcription also detects the language of the recording automatically. As expected, English is the most frequently occurring language; in the US, it accounts for 92.35\% of the calls, an overwhelming majority, followed by Spanish (3.80\%) and Mandarin Chinese (1.30\%). In international calls, English is still the most common language (59.44\%), but other languages occur more frequently than in US calls.

\input{graphics/tables/most_common_languages}

\subsubsection{Similarities Across Different Languages}

Table~\ref{tab:most-frequent-words} shows the top 10 most frequent words occurring in international calls for English, Spanish, and Polish with their translations into English and the normalized word frequency:
\begin{equation}
f_{\text{norm}}(w) = \frac{\text{count}(w)}{\text{count}(w_{\max})}
\label{eq::normalized-frequency}
\end{equation}
where $\text{count}(w)$ represents the total number of occurrences of a given word $w$, and $w_{\max}$ represents the most commonly occurring word. Although not identical, the most frequent words for all three languages are words that can be associated with the context of telephone conversations (\textit{press} and \textit{number}). Words that could be attributed to robocall activity such as \textit{loan}, \textit{payment}, \textit{credit}, \textit{important}, \textit{approved}, and \textit{rate} are present in all three languages.

\begin{figure*}[!ht]
    \centering
    \begin{subfigure}[t]{0.48\textwidth}
        \centering
        \includegraphics[width=\linewidth]{graphics/figures/wordclouds/ny2_word_cloud_en_speech_only_transcripts.pdf}
        \caption{US recordings, English language}
        \label{fig:word-cloud-ny2-en}
    \end{subfigure}
    \hfill
    \begin{subfigure}[t]{0.48\textwidth}
        \centering
        \includegraphics[width=\linewidth]{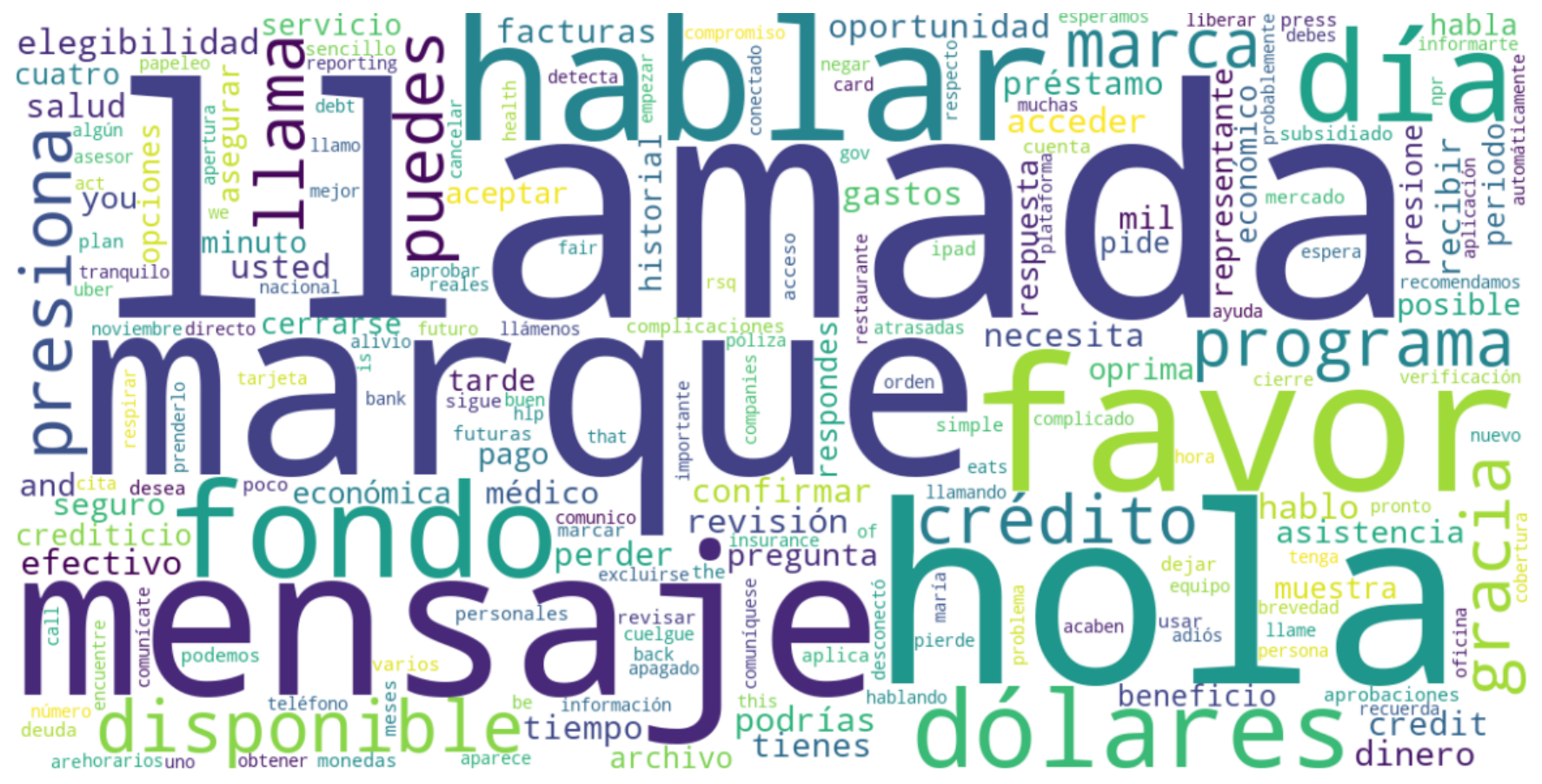}
        \caption{US recordings, Spanish language}
        \label{fig:word-cloud-ny2-es}
    \end{subfigure}
    \vspace*{1.5cm}
    \begin{subfigure}[t]{0.48\textwidth}
        \centering
        \includegraphics[width=\linewidth]{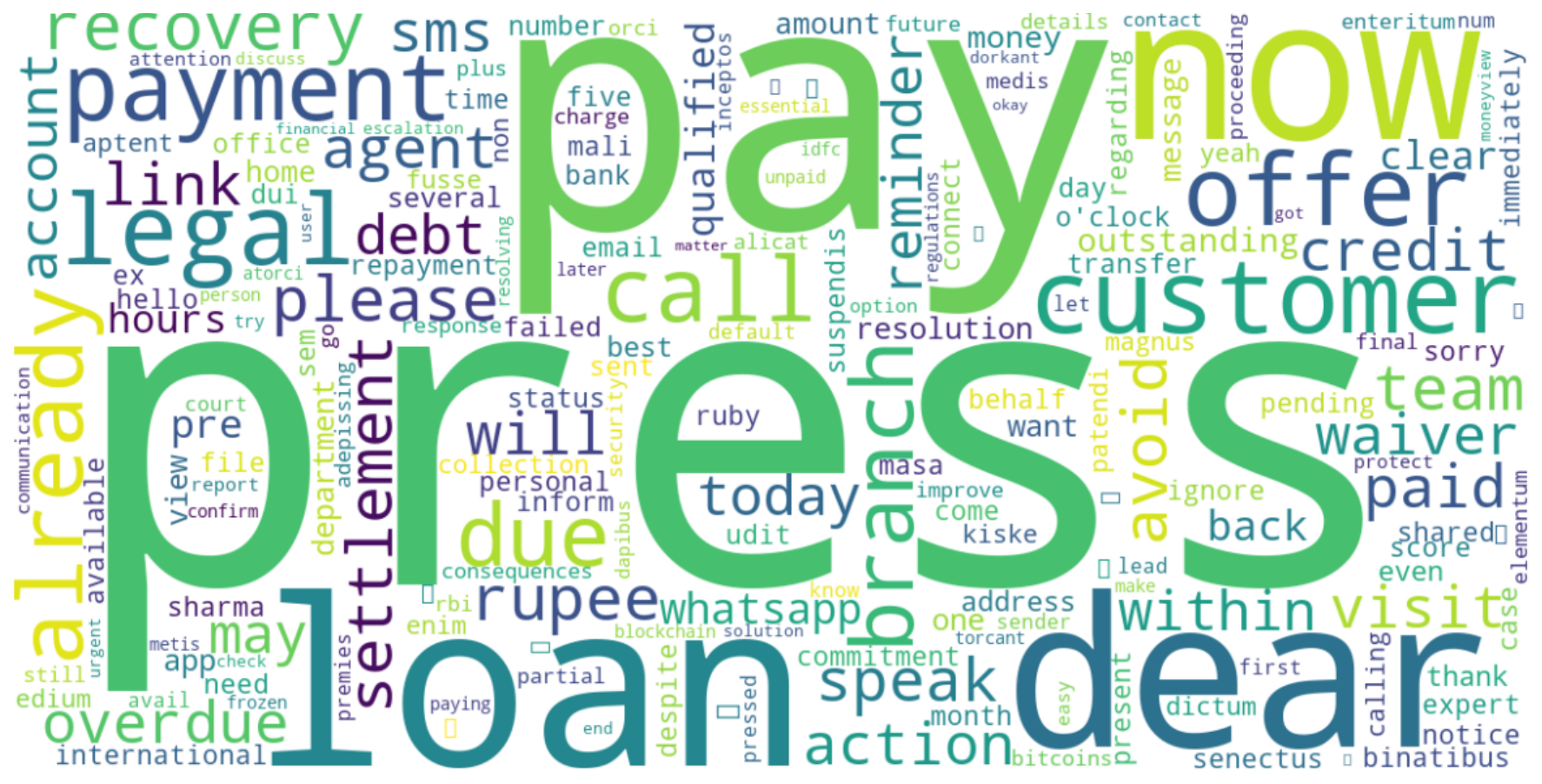}
        \caption{International recordings, English language}
        \label{fig:word-cloud-se1-en}
    \end{subfigure}
    \hfill
    \begin{subfigure}[t]{0.48\textwidth}
        \centering
        \includegraphics[width=\linewidth]{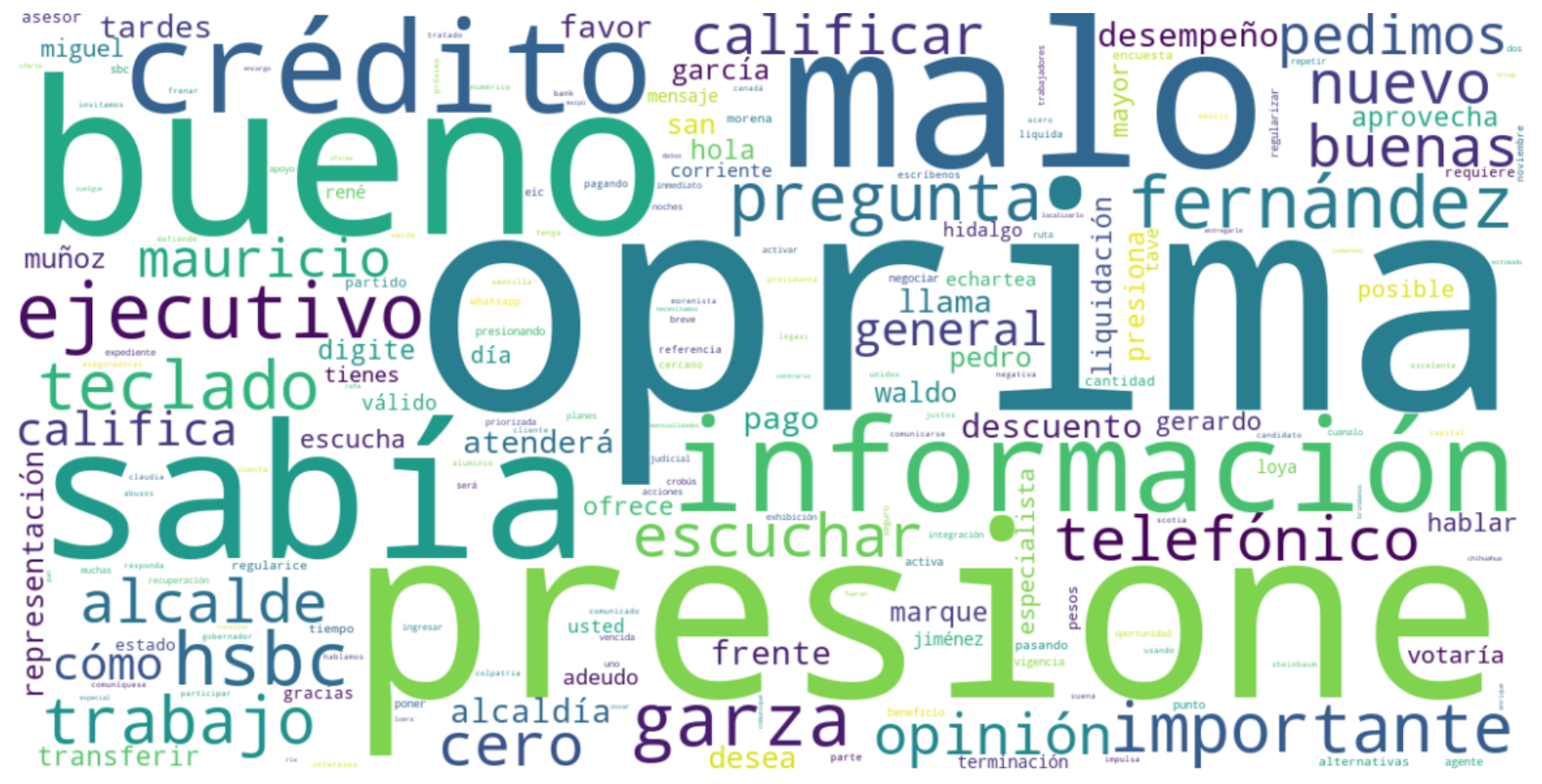}
        \caption{International recordings, Spanish language}
        \label{fig:word-cloud-se1-es}
    \end{subfigure}
    \caption{Word clouds for English and Spanish, generated separately for US recordings and international recordings. Although robocalls were made in the same language, the word clouds differ, indicating that language is not the only predictor of robocall content - the region matters as well.}
    \label{fig:word-clouds-all}
\end{figure*}

\input{graphics/tables/most_frequent_words}

\subsubsection{Regional Differences Between The Same Language}
\label{subsubsec:regional-differences-language}
We want to inspect whether there are differences between the most commonly occurring words of the same language based on the region, i.e. US or international. To illustrate this, we show the word clouds for English and Spanish, per US and international data, in Figure~\ref{fig:word-clouds-all}. Although the exact order and ranking of the words do differ per region, for English the most frequently occurring words are similar in the US and the international recordings, with \textit{press} being the first and second most frequently used word. Words commonly associated with robocalls such as: \textit{loan}, \textit{offer}, \textit{payment}, \textit{now}, \textit{important}, and \textit{financial}, are present in both regions.

One of our observations was that the word \textit{call} is the joint-top occurring word in US calls, but is below the top 10 words in international calls, while the word \textit{press} is the top word in both regions. This pattern is present in Spanish recordings as well. The word \textit{llamada} meaning \textit{call}, and is the most frequent word in US recordings, but is not one of the most common words in international recordings. This may be an indication that robocallers are more likely to use similar scams in the same region, but in different languages, rather than similar scams in the same language but in different regions.

Similarly, the words \textit{oprima} and \textit{presione}, which can both be translated as \textit{press}, are the most frequent words in international recordings. This may indicate that in the US, robocallers are more likely to try to trick the target into calling a specific number, while in international robocalls the goal is to make the victim interact with the robocall. These claims would have to be verified with further analysis.

\subsubsection{Relationship Between Language and Callee Country}

\begin{figure*}[!ht]
  \centering
  \includegraphics[width=0.61\linewidth]{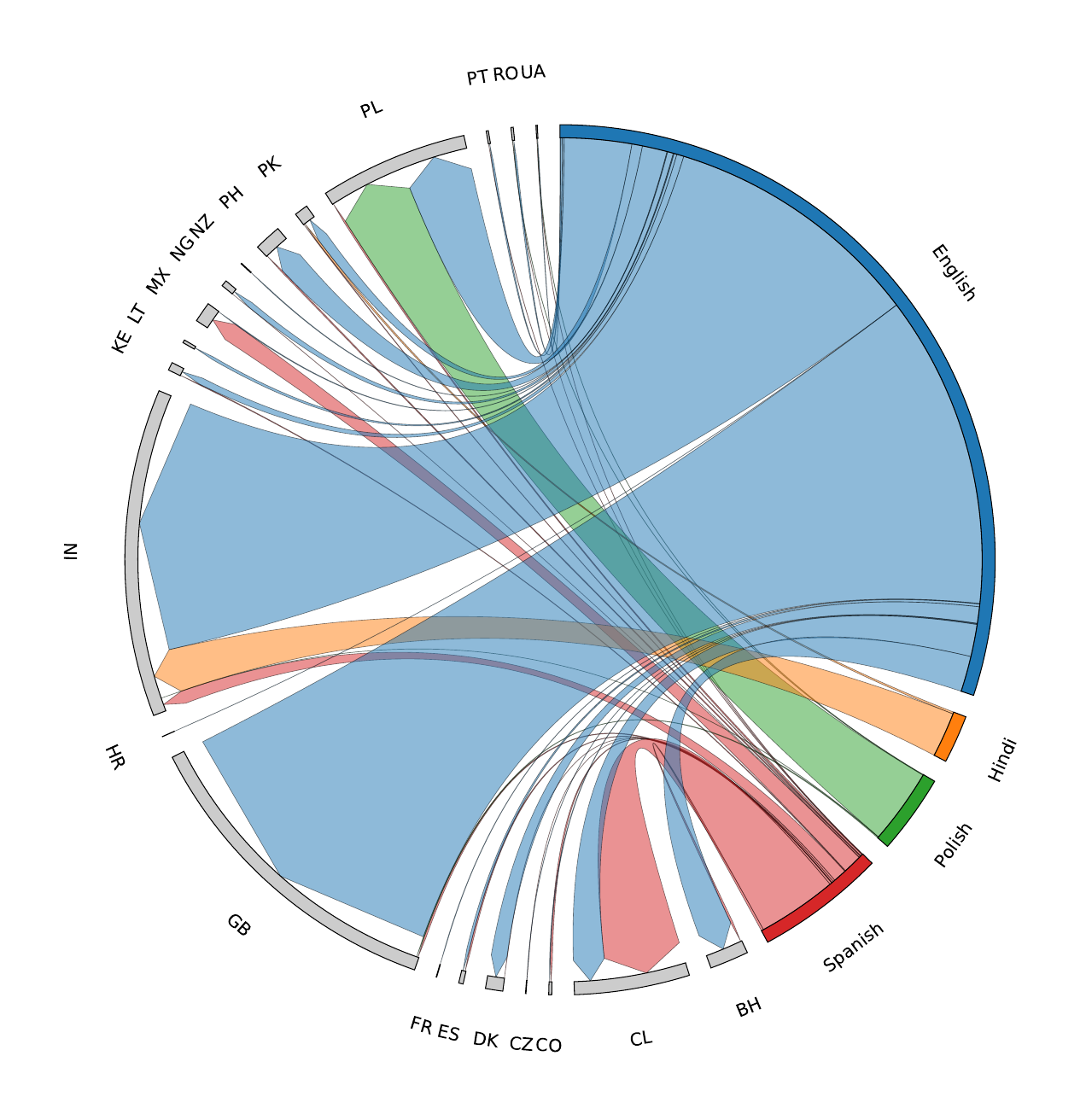}
  \caption{Chord diagram showing the flow of languages towards callee countries. Country names are in ISO 3166-1-alpha-2 codes.}
  \label{fig:chord-diagram-languages}
\end{figure*}

The analysis so far was limited to analyzing the content separately of the caller and callee information. Including the caller number in the analysis can provide some information, but would be of limited reliability since it could be spoofed. We are, however, certain about the callee number, since it belongs to our honeypot. In Figure~\ref{fig:chord-diagram-languages} we present a chord diagram, showing how the recordings, based on language, are distributed among destination countries. An expected result is that English is the most widespread language, with robocalls in English targeting numbers in a majority of countries. Robocalls in Spanish are internationally distributed as well, with calls in Spanish directed towards Bahrain, Chile, India, Mexico, the Philippines, and Poland. Calls in Hindi are directed towards countries in the Indian subcontinent: India and Pakistan. Polish calls are directed overwhelmingly towards Poland, with a small number directed towards the United Kingdom, Romania, and Ukraine, the latter two of which are Slavic countries like Poland. Overall, the language of the robocalls aligns very well with the native languages of the destination countries, while robocalls in English are globally distributed.

\input{graphics/tables/top-clusters-US}
\input{graphics/tables/top-clusters-international}

\subsection{Identifying Robocalls and Campaigns}

In this subsection, we identify robocall campaigns and classify the most common types of robocall scams.

\subsubsection{Campaign Discovery via Transcript Clustering}

Similar to~\cite{prasad2025characterizing}, we apply Density-Based Spatial Clustering of Applications with Noise (DBSCAN)~\cite{ester1996density} to identify robocall campaigns based on their transcripts.

\textbf{Text normalization.} Each transcript undergoes a text normalization process in which we mask sensitive and variable information that should not impact the assignment of a transcript to a given cluster. For example, telephone numbers are replaced with the string ``[phone]'', monetary amounts with ``[amount]'', URLs with ``[url]'', and personal names are replaced with ``[name]''.

\textbf{Feature representation.} Normalized transcripts are represented as TF-IDF vectors over the vocabulary of unigrams and bigrams. We use sublinear term frequency scaling replacing the raw linear term count with a logarithmic scale:

\begin{equation}
\text{tf}(t, d) = 1 + \log(\text{count}(t, d))
\label{eq::tf-sublinear-term-frequency}
\end{equation}

\input{graphics/tables/robocalls_examples_2}
\begin{figure*}[!ht]
  \centering
  \includegraphics[width=0.75\linewidth]{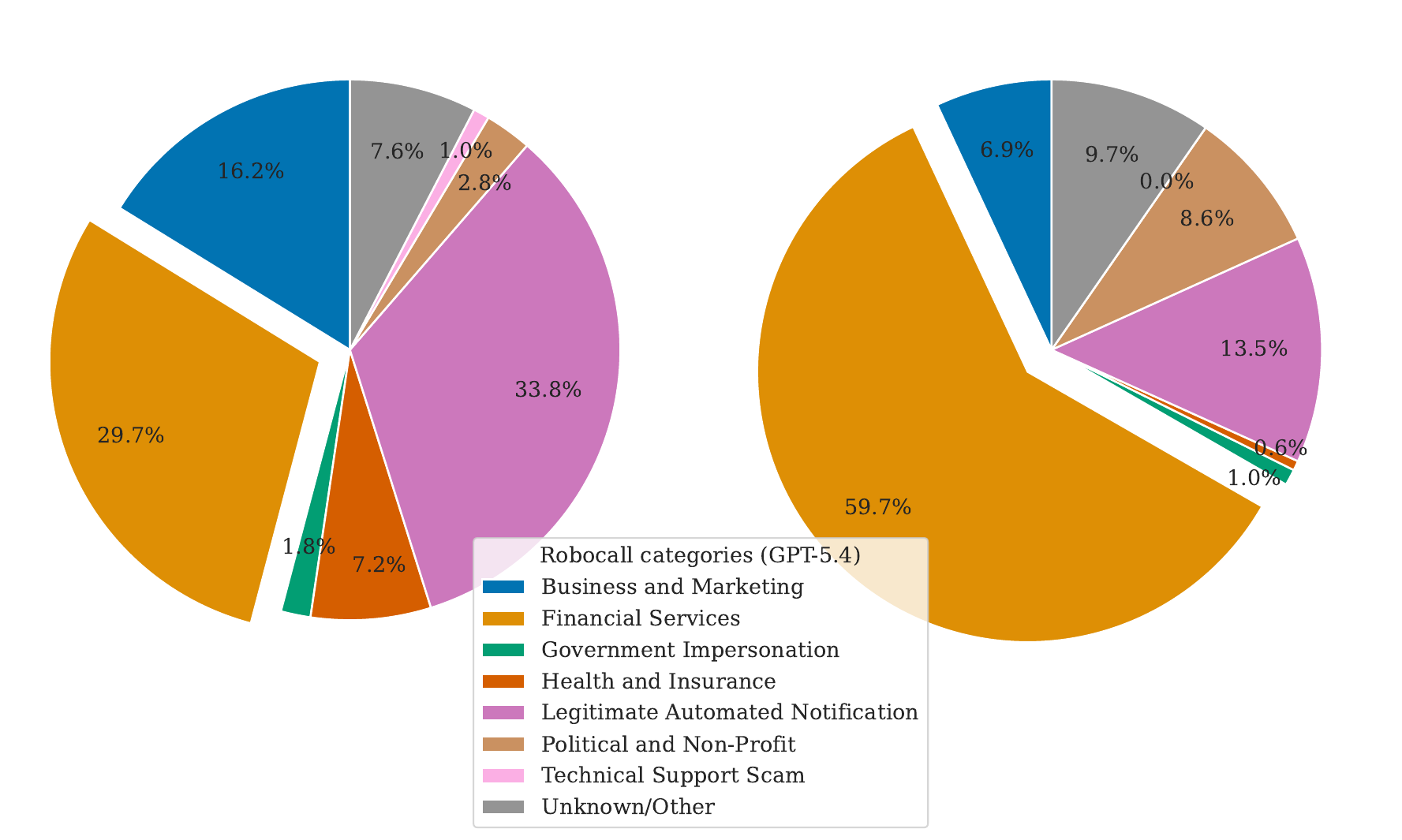}
  \caption{Robocall categorization per region (left - US, right - international). Similar to the results obtained with robocall clusterization (Tables~\ref{tab:top-clusters-us} and~\ref{tab:top-clusters-international}), scams related to financial services are more common in international robocalls, while tech scams are more common in US robocalls. Additionally, US robocalls contain a larger percentage of legitimate automated notifications.}
  \label{fig::pie-chart-robocall-categories}
\end{figure*}

Terms appearing in fewer than 10 transcripts (documents) or in more than 90\% of transcripts are ignored. Transcripts shorter than 40 characters are skipped as well. The vocabulary is capped at \num{15000} features. The resulting sparse vectors are $L2$ normalized, so that the cosine similarity between two transcripts $\mathbf{u}$ and $\mathbf{v}$ is equal to the dot product between them:

\begin{equation}
\cos(\mathbf{u}, \mathbf{v}) = \mathbf{u} \cdot \mathbf{v}
\label{eq::tf-cosine-dot-product}
\end{equation}

\textbf{Clustering.} We use DBSCAN~\cite{ester1996density} to cluster the processed transcripts, represented in the TF-IDF feature space, in an unsupervised manner. DBSCAN has several benefits: (i) it does not require us to specify the number of clusters a priori, (ii) it inherently handles noise and can ignore transcripts that do not belong to any cluster, and (iii) it is agnostic to cluster shape and size. DBSCAN has two parameters which we varied: $minPts \in \lbrace10, 50, 100 \rbrace$ and $\varepsilon \in \lbrace 0.3, 0.4, 0.5, 0.6, 0.7 \rbrace$ by manually inspecting how well the transcripts were clustered.  We finally used the configuration $\varepsilon = 0.4$ for both regions, while $minPts$ was 50 for US transcripts and 10 for international transcripts.

We apply the clustering procedure only to transcripts that have more than 25\% speech content. Combined with filtering transcripts based on their length, this left \num{16177} US transcripts and \num{2701} international transcripts for clustering. A majority of the transcripts are considered to be noise, \num{11693} in the US and \num{1894} internationally. In Tables~\ref{tab:top-clusters-us} and~\ref{tab:top-clusters-international} we show the top 5 clusters of each region.

\textbf{Privacy and availability.} We make publicly available a total of 724 US transcripts and 115 international transcripts which have been verified not to contain any personal information or have been anonymized to conceal first names.

\subsubsection{Few-shot Robocall Categorization}

Large language models (LLMs) have recently emerged as a popular way of processing textual data due to their ability to perform downstream tasks without additional fine-tuning, in a zero-shot or few-shot manner~\cite{brown2020language,wei2022finetuned,sun-etal-2023-text,mu-etal-2024-navigating}. LLMs can be instructed and guided to perform tasks such as classifying audio transcripts as robocalls and categorizing them. We perform few-shot classification: the LLM is provided with a sufficiently detailed description of the problem (classifying robocalls) in its system prompt, and the description contains a few examples of (non)-robocalls. We used ChatGPT-5.4 to classify robocalls based on their transcript.

The LLM categorized \num{11148} (54.8\%) of US robocalls and \num{3021} (74.45\%) of international robocalls. It could classify robocalls into 7 categories: Business and Marketing, Financial Services, Government Impersonation, Health and Insurance, Legitimate Automated Notification, Political and Non-Profit, and Technical Support Scam. The exact distribution of categories per US and international group is shown in Figure~\ref{fig::pie-chart-robocall-categories}. Table~\ref{tab:robocall-categories} shows a typical robocall from each category. The results of using an LLM to directly classify robocalls are aligned with the robocall clustering approach: both approaches showed that tech support scams are more common in the US, with financial services scams being the most dominant type of scam internationally.

\subsubsection{Callback Number Extraction}

A callback number is a phone number that robocallers share with the callee after the call has been answered~\cite{prasad2023diving}. This is a straightforward strategy for robocallers. First, they make a call from a spoofed phone number and impersonate a legitimate organization. Second, during the robocall, they share another phone number with the callee, the callback number, which is not spoofed, since the robocaller's goal is to gain the target's trust and receive phone calls on the callback number. Usually, receiving a call on the callback number implies that the robocaller has already gained some trust from the callee, as the callee decided to call them. The difficulty with uncovering callback numbers is that they only exist in spoken form, i.e. callback numbers do not appear in metadata, they are only spoken during the call.

We performed callback number extraction by processing call transcripts. The percentage of calls that contain a callback number varies significantly depending on whether the call is US or international: 33.6\% of US calls were found to contain a callback number, compared to 1.8\% of international calls. For international calls, we found a total of 51 unique callback numbers, compared to \num{1513} unique callback numbers in US calls. Four out of the top 5 callback numbers in the US were confirmed to be scam robocallers by RoboKiller~\cite{robokiller}, while the fifth was confirmed to belong to a spam/promotional service. This is aligned with our findings from Section~\ref{subsubsec:regional-differences-language}, where we implied that US robocallers are more likely to manipulate the target into calling a specific number, while international robocallers try to deceive the target by making them interact with the robocall.

%% file: graphics/tables/most_common_languages.tex
\begin{table}
    \centering
    \caption{Most commonly occurring languages.}
    \resizebox{0.85\linewidth}{!}{%
\centering
\begin{tabular}{r|r}
\hline
\multicolumn{1}{c|}{US}    & \multicolumn{1}{c}{International}  \\ 
\hline
English (92.35\%)          & English (59.44\%)                  \\
Spanish (3.80\%)            & Spanish (13.85\%)                  \\
Mandarin Chinese (1.30\%)           & Polish (8.23\%)                    \\
Korean (0.71\%)~ & Hindi (7.10\%)                     \\
French (0.43\%)         & Czech (1.82\%)                   
\end{tabular}
    }
    \label{tab:most-common-languages}
\end{table}

%% file: graphics/tables/most_frequent_words.tex
\begin{table}
    \centering
    \caption{Most frequent words among English, Spanish, and Polish international recordings.}
    \resizebox{\linewidth}{!}{%
\centering
\begin{tabular}{r|r|r}
\hline
\multicolumn{1}{c|}{English} & \multicolumn{1}{c|}{Spanish}     & \multicolumn{1}{c}{Polish}      \\ 
\hline
Press (1.00)                 & Oprima (press) (1.00)            & Sie (myself) (1.00)             \\
Pay (0.58)                   & Presione (press) (0.55)          & Numer (number) (1.00)           \\
Loan (0.42)                  & Bueno (good) (0.43)              & Prosze (please) (0.47)          \\ 
\hline
Dear (0.34)                  & Malo (bad) (0.38)                & Dzien (day) (0.46)              \\
Now (0.31)                   & Sabia (know) (0.32)              & Dobry (good) (0.45)             \\
Customer (0.26)              & Informacion (information) (0.29) & Zostalo (remaining) (0.33)      \\ 
\hline
Legal (0.24)                 & Credito (credit) (0.28)          & Online (online) (0.32)          \\
Offer (0.23)                 & Ejecutivo (executive) (0.26)     & Zatwierdzone (approved) (0.32)  \\
Payment (0.21)               & HSBC (HSBC) (0.25)               & WhatsApp (WhatsApp) (0.32)      \\
Call (0.20)                  & Importante (0.24)                & Stawka (rate) (0.32)            
\end{tabular}
    }
    \label{tab:most-frequent-words}
\end{table}

%% file: graphics/tables/top-clusters-US.tex
\begin{table*}
    \centering
    \caption{Top 5 discovered US robocall clusters. The most common type of robocalls are tech scams, most notably impersonation of the company Google.}
    \resizebox{0.75\linewidth}{!}{%
\centering
\begin{tabular}{l|r|r} 
\hline
\multicolumn{1}{c}{Cluster}                                                & \multicolumn{1}{c}{Count} & \multicolumn{1}{c}{Top words}                             \\ 
\hline
Benefits/Grants Scam                                                       & 460 (10.26\%)             & benefits, period, program, now, press, case               \\
\hline
Google Business Listing Scam     & 245 (5.46\%)              & google, voice, business, press, search, customers         \\
\hline
Google Business Listing Scam (II) & 239 (5.33\%)              & google, agent, voice, business                            \\
\hline
Exclusive Benefits Spam                                                    & 204 (4.55\%)              & benefits, exclusive, thousands, quickly, pocket           \\
\hline
Caller ID Spoofing                                                         & 196 (4.37\%)              & return, assist, urgent, time-sensitive, capital recovery 
\end{tabular}
    }
    \label{tab:top-clusters-us}
\end{table*}

%% file: graphics/tables/top-clusters-international.tex
\begin{table*}
    \centering
    \caption{Top 5 international robocall clusters. The most common scam types are related to loan/debt scams, with 3 out of the top 5 clusters belonging to this type of scam. Scams are multilingual: English, Hindi, and Spanish.}
    \resizebox{0.65\linewidth}{!}{%
\centering
\begin{tabular}{l|r|r} 
\hline
\multicolumn{1}{c}{Cluster}                                       & \multicolumn{1}{c}{Count} & \multicolumn{1}{c}{Top words}                  \\ 
\hline
Loan/Debt Scam (Hindi)  & 83 (10.29\%)              & legal, recovery, notice, final, agent          \\
\hline
Loan/Debt Scam (II) (English) & 59 (7.31\%)               & qualified, loan, prequalified, Whatsapp, link  \\
\hline
PayMe Loan (English)     & 43 (5.33\%)               & PayMe, loan, reminder, visit, delay            \\
\hline
Political Spam (Spanish) & 40 (4.96\%)               & important, mayor, performance, opinion         \\
\hline
Loan/Debt Scam (III) (Hindi) & 35 (4.34\%)               & loan, overdue, legal, notice, visit           
\end{tabular}
    }
    \label{tab:top-clusters-international}
\end{table*}

%% file: graphics/tables/robocalls_examples_2.tex
\begin{table*}
    \centering
    \caption{Extracted robocall categories and examples of typical robocall scripts from each category. A quick internet search on the callback numbers from the transcripts reveal that these numbers are well-known robocallers with bad caller reputation.}
    \resizebox{1.0\linewidth}{!}{%
\centering
\begin{tabular}{c|l}
\hline
Category                          & \multicolumn{1}{c}{Example}                                                                                                                                                                                                                                                                                                                                                                                                                                                                                                                                                                                                                                                                                                                                                                                                                                                                                                                                                                                  \\ \hline
Business and Marketing            & \begin{tabular}[c]{@{}l@{}}Hello, your timeshare details just came across my desk and there is something you should know.\\ Call me back ASAP 949-237-7769. Again, that phone number is 949-237-7769. Thank you. Talk to you soon.\end{tabular}                                                                                                                                                                                                                                                                                                                                                                                                                                                                                                                                                                                                                                                                                                                                                              \\ \hline
Financial Services                & \begin{tabular}[c]{@{}l@{}}Hello, this is Personal Loan Locator, just giving me a real quick call in regards to the recent personal loan request you \\ submitted. I have a couple alternative options for you, as well as some additional resources to go over. So go ahead and\\give me a call back at 855-926-6692. Again, that phone number was 855-926-6692. To opt out, go ahead and press 9. Thank you.\end{tabular}                                                                                                                                                                                                                                                                                                                                                                                                                                                                                                                                                                                \\ \hline
Government Impersonation          & \begin{tabular}[c]{@{}l@{}}This is Victoria Masterson with Pre-Legal Services. This is our final attempt to notify you regarding legal documents that\\ have been returned. Please contact my office at 301-298-1127. Again, that's 301-298-1127. Thank you.\end{tabular}                                                                                                                                                                                                                                                                                                                                                                                                                                                                                                                                                                                                                                                                                                                                  \\ \hline
Health and Insurance              & \begin{tabular}[c]{@{}l@{}}Eligibility Patient Advocacy Liaisons is calling Samia to provide a free service regarding a personal matter. This is a free service.\\ This is not a sales or collections call. Please call ePALS back at 346-738-0894 and quote reference number account number.\\ Again, our number is 346-738-0894 and quote reference number account number. Thank you.\\ Enlaces de apoyo para el paciente lo es de la mando a Samia para ofrecer un servicio gratuito con respecto a un asunto personal,\\ no es una la manda de ventas o cabranzas. Favor de la mando enlaces de apoyo para el paciente al telefono 346-738-0894\\ con el número de referencia account number. De nuevo, el número de teléfono es 346-738-0894 y el número de\\ referencia es account number. Gracias.\end{tabular}                                                                                                                                                                                       \\ \hline
\begin{tabular}[c]{@{}c@{}}Legitimate Automated\\ Notification\end{tabular} & \begin{tabular}[c]{@{}l@{}}This message is in English and Espanol. This is a message from the Department of Public Works. Please note there is a household\\ hazardous waste drop-off day this Saturday, October 25th from 9 a.m. to 12 p.m. at the Wastewater Treatment Plant, 40 South\\ Porter Street. Proof of Haverhill Residency is required, and there is a 25-pound-slash-25-gallon limit per vehicle.\\ Este es un mensaje del Departamento de Obras Públicas. Tenga en cuenta que hay un día de entrega de desechos domésticos\\ peligrosos este sábado 25 de octubre de 9 a.m. a 12 p.m. en la planta de tratamiento de aguas residuales, 40 South Porter Street.\\ Se requiere prueba de residencia en Averillie y hay un límite de 25 libras, barra 25 galones por vehículos.\end{tabular}                                                                                                                                                                                                 \\ \hline
Political and Non-Profit          & \begin{tabular}[c]{@{}l@{}}Greetings, neighbors. This is your 16th Ward Alderman Stephanie Coleman, and I warmly invite you and your family\\ to the 16th Ward monthly community meeting and turkey giveaway on Saturday, November 8th. Join us at a special\\ location for this event only, the Salvation Army Red Shield Adele Center, located 949 West 69th Street from\\ 10 a.m. until 12 noon. We will have representatives from the Chicago Department of Public Health, Howard Brown\\ Health, and the Greater Chicago Food Depository to share valuable information and resources available for our community,\\ along with a light holiday meal to enjoy together. Be a part of the change that's happening in the 16th Ward as we build \\ stronger communities together. Mark your calendars for Saturday, November 8th at a special location, Salvation Army\\ Red Shield Adele Center at 949 West 69th Street from 10 a.m. until 12 noon. Don't just meet me there, beat me there.\end{tabular} \\ \hline
Technical Support Scam            & \begin{tabular}[c]{@{}l@{}}This call is inform you that your Amazon Prime subscription has been auto-renewed and \$499 has been debited from\\ your account successfully. If you want to cancel this subscription, please press 1 to talk to our executive.\end{tabular}                                                                                                                                                                                                                                                                                                                                                                                                                                                                                                                                                                                                                                                                                                                                    
\end{tabular}
    }
    \label{tab:robocall-categories}
\end{table*}

%% file: text/discussion.tex
\section{Discussion}

Even though robocalls are an international problem, our findings indicate that US citizens are substantially more affected than the rest of the world. The median number of phone numbers in our US honeypot was \num{11751}, compared to \num{101913} international phone numbers. Despite an almost tenfold difference in the number of available phone lines, the US honeypot numbers received \num{8314813} calls (95.1\%) compared to \num{432538} calls (4.9\%) received by the international numbers. On average, a single phone number in our US honeypot received 707.5 calls during a nine-month period, about 2.62 calls per day. Non-US numbers received, on average 4.24 calls during the same period, about 0.016 calls per day.

Despite the difference in frequency, robocalls made towards US and international citizens share many similarities. We showed that the temporal calling patterns are alike: robocalls tend to be made during working hours, with more than 60\% of robocalls made between 9AM and 5PM. We argue that this makes sense for malicious robocallers: they want to catch the callee off guard. The distribution of calls is heavily skewed in both regions: usually, a few callers account for a majority of the calls, with the top 1\% of callers accounting for 24.4\% of all calls.

Geographically, both US and international phone calls share the same characteristics: 89.9\% of calls display a caller ID with the country matching the callee country. Since caller IDs are not fully reliable, we tested whether call timing and shared callees reveal structure that is also supported by similar call content. In both regions, caller IDs that reach the same callee within short time windows also tend to deliver similar messages.

Furthermore, call content analysis revealed that a significant number of robocalls are silent, which is aligned with previous research findings. We showed that performing automatic speech recognition on call recordings should be done with caution, as hallucinations can yield unreliable transcripts. When changing the operating mode of our honeypot from passive to interfering, we noticed significant drops in call volume. This may be an indication that robocallers are monitoring who or what is answering the call and refrain from playing any messages unless they are certain that a human has answered.

English is by far the most common language in all recorded calls, followed by Spanish, Mandarin Chinese, Polish, and Hindi. Analyzing the most frequently occurring words across all languages, we observed a pattern of words that indicates typical robocall behavior: triggering a sense of urgency and terms related to business, payment, debt, loan, and tech support. When analyzing calls in the same language but a different region (e.g. Spanish or English in US and international calls), we observed a difference between the top words: US calls persuade the callee to call another number, while international calls tend to be more interactive (press a button). This is further supported by our callback number extraction, where 33.6\% of US calls were found to contain a callback number, compared to just 1.8\% of international calls.

Finally, we used two approaches to cluster and categorize robocalls. In an unsupervised approach using DBSCAN, we uncovered that top US clusters are related to business and tech scams, while top international clusters are loan and debt scams. With few-shot classification, we used an LLM to categorize robocalls and obtained similar results: 59.7\% of international robocalls were classified as related to financial services compared to 29.7\% of US robocalls, 16.2\% of US calls were categorized as related to business and marketing, compared to 6.9\% of international calls.

\subsection{Limitations}

Working with robocalls is a sensitive topic, from both a legal and ethical perspective. We tried our best to preserve caller privacy and ensure no legitimate calls were processed during our research. This, however, resulted in our study introducing several limitations, such as playing a warning message whenever calls were answered or omitting caller IDs that made only a single call to the honeypot. Similarly, the number of transcripts and audio recordings released in the dataset is below \num{2000}, as they require manual confirmation by human annotators, often in multiple languages. The quantity of audio recordings and transcripts in the dataset is therefore relatively small compared to the 8.7 million CDRs that are released without requiring human annotation.

\subsection{Remedies and Future Directions}

It would be interesting to obtain legal permission to spoof/fake answered calls, instead of playing warning messages. By doing so, we could analyze whether the number of received phone calls would still drop once we start answering them.

It would also be interesting to repeat the experiments without some of the technical limitations mentioned in Section~\ref{section::method}. For example, it would be beneficial to split the international honeypot into smaller groups or regions, instead of treating them as a single group. This could provide more detailed insights into per-country statistics.

Finally, by collaborating with government institutions and telecoms, it may be possible to record and make publicly available a larger collection of robocalls.

For future work, we intend to extend our dataset and to investigate whether any of the limitations of this study can be addressed. We plan to work on systems that can detect robocalls in real time, regardless of the language or country of the callee, and to make such solutions publicly available.

%% file: text/conclusion.tex
\section{Conclusion}

In this paper, we presented a detailed overview of the state of international robocalls. Our honeypot registered 9.6 million calls, of which we make publicly available a dataset of 8.7 million call detail records, 677 robocall recordings, and 839 clustered and categorized robocall transcripts. Our analysis showed that although robocalls are an issue in many countries, the threat severity is highest in the US. We analyze temporal and geographical calling patterns and examined co-targeting behavior within short time windows. By analyzing call content, we showed the most frequently occurring languages in robocalls and their (dis)similarities based on region. We used an unsupervised clustering approach to identify frequent robocall campaigns, as well as an LLM-based approach that categorized robocalls into common categories via few-shot prompting. For future work, we plan to extend our dataset and overcome some of the technical limitations of our study, as well as to work on directly combating robocalls.

\section{Acknowledgments}
This research was supported in part by the project Infobip Global Communication Platform (PK.1.1.07.0001), part of the Important Project of Common European Interest on Next Generation Cloud Infrastructure and Services (IPCEI-CIS) consortium.
The authors also want to thank Patricio Marcos Petrić for his assistance during the writing of this paper and support in robocall analysis, as well as Jovi Teoh, Kyung Hyun Nam, and Bartosz Nabrdalnik for their time and assistance in analyzing robocalls in different languages.